\documentclass[fleqn,usenatbib]{mnras}

\usepackage[T1]{fontenc}

\DeclareRobustCommand{\VAN}[3]{#2}
\let\VANthebibliography\thebibliography
\def\thebibliography{\DeclareRobustCommand{\VAN}[3]{##3}\VANthebibliography}

\usepackage{graphicx}	
\usepackage{amsmath}	
\usepackage{amssymb}	
\usepackage{adjustbox}
\usepackage{makecell}
\usepackage{siunitx}
\usepackage{hyperref}
\usepackage[utf8]{inputenc}
\usepackage{MnSymbol}

\usepackage{newtxtext,newtxmath}

\title[Circumnuclear Molecular Gas Reservoirs]{WISDOM Project - XVI. The link between circumnuclear molecular gas reservoirs and active galactic nucleus fuelling}

\author[J. S. Elford et al.]{\parbox{\textwidth}{
Jacob S. Elford,$^{1}$\thanks{E-mail: elfordj@cardiff.ac.uk} 
Timothy A. Davis,$^{1}$ 
Ilaria Ruffa,$^{1}$
Martin Bureau,$^{2}$
Michele Cappellari,$^{2}$
Jindra Gensior,$^{3}$ 
Satoru Iguchi,$^{4,5}$ 
Fu-Heng Liang,$^{2}$
Lijie Liu,$^{6,7}$
Anan Lu,$^{8}$ 
Thomas G. Williams$^{2}$}\vspace{0.3cm}
\\ 
$^{1}$Cardiff Hub for Astrophysics Research \&\ Technology, School of Physics \&\ Astronomy, Cardiff University, Queens Buildings, The Parade, Cardiff, CF24 3AA, UK
\\
$^{2}$Sub-department of Astrophysics, Department of Physics, University of Oxford, Denys Wilkinson Building, Keble Road, Oxford OX1 3RH, UK
\\
$^{3}$Institute for Computational Science, University of Z\"{u}rich, Winterthurerstrasse 190,8057 Z\"{u}rich
\\
$^{4}$Department of Astronomical Science, SOKENDAI (The Graduate University of Advanced Studies), Mitaka, Tokyo 181-8588, Japan
\\
$^{5}$National Astronomical Observatory of Japan, National Institutes of Natural Sciences, Mitaka, Tokyo 181-8588, Japan
\\
$^{6}$Cosmic Dawn Center (DAWN)
\\
$^{7}$DTU-Space, Technical University of Denmark, Elektrovej 327, DK-2800 Kgs, Lyngby , Denmark
\\
$^{8}$McGill Space Institute and Department of Physics, McGill University, 3600 rue University, Montreal, QC H3A 2T8, Canada
}

\date{Accepted XXX. Received YYY; in original form ZZZ}

\pubyear{2021}

\begin{document}
\label{firstpage}
\pagerange{\pageref{firstpage}--\pageref{lastpage}}
\maketitle

\begin{abstract}
We use high-resolution data from the millimetre-Wave Interferometric Survey of Dark Object Masses (WISDOM) project to investigate the connection between circumnuclear gas reservoirs and nuclear activity in a sample of nearby galaxies. Our sample spans a wide range of nuclear activity types including radio galaxies, Seyfert galaxies, low-luminosity active galactic nuclei (AGN) and inactive galaxies.
We use measurements of nuclear millimetre continuum emission along with other archival tracers of AGN accretion/activity to investigate previous claims that at, circumnuclear scales (<100\,pc), these should correlate with the mass of the cold molecular gas. 
We find that the molecular gas mass does not correlate with any tracer of nuclear activity. This suggests 
the level of nuclear activity cannot solely be regulated by the amount of cold gas around the supermassive black hole (SMBH). This indicates that AGN fuelling, that drives gas from the large scale galaxy to the nuclear regions, is not a ubiquitous process and may vary between AGN type, with timescale variations likely to be very important. 
By studying the structure of the central molecular gas reservoirs, we find our galaxies have a range of nuclear molecular gas concentrations. This could indicate that some of our galaxies may have had their circumnuclear regions impacted by AGN feedback, even though they currently have low nuclear activity. On the other hand, the nuclear molecular gas concentrations in our galaxies could instead be set by secular processes. 
\end{abstract}




\section{Introduction}\label{sec:Section1}
It has been established that a supermassive black hole (SMBH) exists at the centre of almost all massive galaxies ($M_{*}\gtrsim10^{9.5}\,\rm M_{\odot}$). A large number of studies have shown that tight correlations exist between the masses of such SMBHs and the properties of their host galaxies (such as the bulge mass: \citealp[e.g.][]{1998AJ....115.2285M, 2003ApJ...589L..21M} and velocity dispersion: \citealp[e.g.][]{2000ApJ...539L...9F,2002ApJ...574..740T,2009ApJ...698..198G}), suggesting a self-regulated co-evolution between them (see e.g.\,\citealt{2013ARA&A..51..511K} for a review). 
There is evidence that active galactic nuclei (AGN) and the associated energetic output can play a crucial role in setting up and maintaining SMBH-host galaxy co-evolution, as it can change the physical conditions of the surrounding interstellar medium (ISM) and/or expel it from the nuclear regions \citep[e.g.][]{2006MNRAS.370..645B,2006MNRAS.365...11C,2015ARA&A..53..115K,2017FrASS...4...42M,2017NatAs...1E.165H}. The many details of these processes, however, are still poorly understood. 
\par
In the local Universe (\textit{z}\,<\,0.1), the AGN population can be separated into two main (non-exclusive) groups, differentiated by the mode of dominant energetic output: radiative and kinetic \citep[e.g.][]{2014ARA&A..52..589H}. In the former the accretion occurs at high rates ($\gtrsim1$\% of the Eddington limit) through optically-thick and geometrically-thin discs \citep{1973A&A....24..337S}. This mode is radiatively efficient, so the dominant energy output is from the conversion of the potential energy of the matter accreted onto each SMBH into electromagnetic radiation. Kinetic-mode AGN instead produce  little radiation and channel the bulk of the energy generated from the accretion process into collimated outflows of non-thermal plasma (i.e.\,radio jets). In these objects the geometrically-thin accretion disc is absent or truncated at some inner radii and likely replaced by geometrically-thick, optically-thin advection-dominated accretion flows \citep[i.e.\,ADAFs;][]{1995ApJ...452..710N}, whereby the material is accreted onto the SMBH at low rates ($\ll1$\% of the Eddington limit).
\par
Nearby radiative-mode AGN with weak or no radio jet emission have historically been called Seyfert galaxies. These objects possess all the characteristics of the “conventional'' AGN described in the framework of the standard unified model \citep[e.g.][]{Antonucci93,UrryPadovani95}, and are typically hosted by late Hubble type galaxies \citep[e.g.][]{Martini03}. AGN producing strong kinetic feedback are instead typically identified as radio galaxies (RGs) and - based on their optical spectra - can be divided into two main classes \citep[e.g.][]{2012MNRAS.421.1569B}: high-excitation radio galaxies (HERGs) and low-excitation radio galaxies (LERGs). The former show strong high-ionisation (Seyfert-like) emission lines in their optical spectra, produce both radiative and kinetic AGN feedback, and are typically hosted by massive ($M_{\filledstar}\gtrsim10^{9.5}\,M_{\odot}$) early-type galaxies (ETGs). LERGs show no or weak, LINER (low-ionisation nuclear emission-line region)-like emission lines in their optical spectra, produce almost exclusively kinetic feedback, and are typically found in very massive ($M_{\filledstar}\gtrsim10^{11}\,M_{\odot}$) ETGs. Understanding the source of SMBH fuelling in the AGN populations introduced above is crucial to putting constraints on the physical processes driving and regulating the SMBH-host galaxy co-evolution. Up to now, however, a general picture for the fuelling of active SMBHs in the local Universe is still missing.
\par
The central regions of Seyfert galaxies have been often observed to be dominated by cold atomic and molecular gas  \citep[e.g.][]{2013A&A...558A.124C,2014A&A...567A.125G}, suggesting a potential connection with their nuclear activities. The finding by \citet{Izumi16} of a positive correlation between the mass of $\approx100$pc-scale circumnuclear disks (CNDs) of dense molecular gas and the black hole mass accretion rate in nearby Seyferts seems to support this hypothesis. More recently, \citet{2021A&A...652A..98G} also found that nuclear activity in these objects can cause deficits in their circumnuclear molecular gas reservoirs, with a negative trend between nuclear 2-10~keV X-ray luminosity and the central gas concentration. In both cases, however, the studies have been conducted on small samples of about 10 {AGN of the same type}. These samples also span relatively small ranges of AGN luminosities ($L_{\rm 2-10keV}\sim10^{41}-10^{44}$~erg~s$^{-1}$) and host galaxy properties (almost exclusively barred spirals). It is therefore currently not clear whether or not the inferred cold gas-nuclear activity connection would hold over a broader population of active galaxies.

On the other hand, a long-established scenario suggests that the HERG/LERG dichotomy may be a consequence of different sources for the accreting gas. 
In this framework, HERGs are fuelled at relatively high rates by cold gas acquired from merging or collisions with gas-rich galaxies \citep[e.g.][]{2012MNRAS.421.1569B}. LERGs 
are instead powered by the accretion of hot gas from the intergalactic medium (IGM) through Bondi spherical accretion \citep{1952MNRAS.112..195B,2007MNRAS.376.1849H}. This hypothesis was initially supported by studies finding a correlation between jet power and Bondi accretion rate in LERGs \citep[e.g][]{2006MNRAS.372...21A, 2007MNRAS.376.1849H,2008A&A...486..119B}. Over the past decade, however, strong evidence has been acquired that cold gas can also play a role in fuelling LERGs, as large masses of cold gas and dust have been often observed at the centres of these objects \citep[i.e.\,$M_{\rm H_2} \sim 10^{7} - 10^{10}$~M$_{\rm \odot}$;][]{2010A&A...523A..38P, 2010A&A...518A...9O, North19,2019MNRAS.484.4239R,2019MNRAS.489.3739R,Ruffa20}. The {\em total} molecular gas mass of a sample of nearby ETGs (most of which are LERG hosts) have also been observed to {weakly} correlate with the AGN jet power, providing further evidence that there could be a close connection between the two \citep{Babyk19}. Models for cold gas SMBH fuelling in typical LERG hosts have been also developed and imply that the observed cold gas reservoirs originate from cooling of the hot X-ray emitting surrounding halos, either directly and smoothly \citep[e.g.][]{Negri14} or after chaotic cooling  (as predicted in chaotic cold accretion models, CCA; \citealp[e.g.][]{2007MNRAS.377L..25K, 2009ApJ...702...63W, 2012ApJ...753...15N, 2013MNRAS.432.3401G, 2015A&A...579A..62G, 2017MNRAS.466..677G, 2015MNRAS.453L..46K}). Growing observational evidence provide support to this picture, at least for LERGs located in high-density environments (i.e.\,in rich groups and clusters). The importance of (chaotic) hot gas cooling in more isolated LERGs is still not clear \citep[e.g.][]{2019MNRAS.489.3739R,2022MNRAS.510.4485R,Maccagni23}. 

In general, both theoretical studies \citep[e.g.][]{1989Natur.338...45S} and numerical simulations \citep[e.g.][]{2005ApJ...632..821P,2010MNRAS.408..961P,2012ApJ...757..136W,2016ApJ...830...79M} have shown that cold gas can play a fundamental role in fuelling nearby AGN (both radiative and kinetic mode), with \cite{2022MNRAS.514.2936W} finding that AGN are preferentially located in galaxies with high molecular gas fractions. A corresponding comprehensive observational picture, however, is still missing.
\par
\par
The mm-Wave Interferometric Survey of Dark Object Masses (WISDOM) project is exploiting high-resolution CO observations from the Atacama Large Millimeter/submillimeter Array (ALMA) with the primary aim of measuring SMBH masses in a morphologically-diverse sample of nearby galaxies \citep[e.g.][]{2017MNRAS.468.4663O, 2017MNRAS.468.4675D, 2018MNRAS.473.3818D, 2019MNRAS.485.4359S, North19, 2021MNRAS.500.1933S, 2021MNRAS.503.5984S, 2021MNRAS.503.5179N, 2022MNRAS.516.4066L,2023MNRAS.522.6170R}. In this paper, we use WISDOM data with a typical spatial resolution of $\sim$20--30\,pc to look for a connection between the \textit{circumnuclear} molecular gas reservoirs observed with ALMA and the SMBH fuelling across a sample with a { wide} range of nuclear activities (from low/high luminosity Seyferts to LERGs). Our main aim is to explore the scenarios described above, testing the cold gas-SMBH fuelling correlations and the scales over which it persists. 
\par
This paper is organised as follows. In Section \ref{Section 2}, we describe the sample and the multi-wavelength observations used for our analysis. We describe the adopted methodology in \ref{Section 3}. We present our results in Section \ref{Section 4} and discuss them in Section \ref{Section 5}, before summarising and concluding in Section \ref{Section 6}.

\section{Observations}
\label{Section 2}
\subsection{WISDOM sample}
\label{Section 2.1}
WISDOM ALMA data (with a typical resolution of $\approx0\farcs 1$ or 30\,pc) were originally collected with the intent of measuring SMBH masses. The main selection criterion for WISDOM galaxies was thus to have the SMBH sphere of influence (SOI)\footnote{The SOI is the region where the gravitational potential of the SMBH dominates over that of the host galaxy and is defined as 
\begin{equation}
R_{\rm SOI}\equiv G M_{\rm BH}/\sigma_*^2,   
\end{equation}
where $M_{\rm BH}$ is the mass of the SMBH, $\sigma_*$ is the stellar velocity dispersion of the host bulge and \textit{G} is the gravitational constant.} spatially-resolvable with ALMA.  
Therefore, our sample of galaxies is fairly heterogeneous, containing both nearly-quenched ETGs and star-forming spirals with a range of nuclear activities. In particular, here we study data of 35 WISDOM objects, spanning stellar masses $M_{\filledstar}$ from $10^{9.1}$ to $10^{11.8} \ \rm M_{\odot}$, and 1.4~GHz radio luminosities $L_{\rm 1.4GHz}$ from \, $\approx \rm 10^{34}$ to $\approx \rm 10^{41} \ ergs \ s^{-1}$. 
The sample galaxies and their basic parameters are listed in Table \ref{tab:Table 1}. The AGN properties of our sample sources are discussed in detail below, and comparisons of these with other literature samples are presented in Section \ref{samplefigs}.

\begin{table*}
\centering
\caption{Physical parameters of the galaxies sample.}
\begin{tabular}{l
S[table-format=3.3]
c
c
c
c
c
cccc}
\hline
{Galaxy} & {Distance} & {Jet} & {Galaxy type} & {AGN type} & {log($ M_\filledstar/{\rm M}_{\odot}$)} & {Mass Ref} & {log(SFR/$\rm M_{\odot} \, yr^{-1}$)} & {Project code} & {Reference}\\
& {(Mpc)} & & & & & & & &\\
{(1)} & {(2)} & {(3)} & {(4)} & {(5)} & {(6)} & {(7)} & {(8)} & {(9)} &{(10)}\\
\hline
FRL49 & 85.7 & No & E-S0 & Seyfert 2 & 10.30 & L22 & 0.78 & b & \cite{2022MNRAS.516.4066L} \\
FRL1146 & 136.7 & Yes & Sc & Seyfert 1 & 11.32 & ${\rm M}_{K_s}$ & - & a,b & This work \\
MRK567 & 140.6 & No & Sc & - & 11.26 & C17 & 1.30 & a,b & \cite{2022MNRAS.512.1522D} \\
NGC0383 & 66.6 & Yes & E-S0 & LERG & 11.82 & MASSIVE & 0.00 & c,d,e & \cite{North19} \\
NGC0404 & 3.0 & No & E-S0 & LINER & 9.10 & S10 & -3.04 & f & \cite{2020MNRAS.496.4061D} \\
NGC0449 & 66.3 & No & SBa & Seyfert 2 & 10.07 & z0MGS & 1.19 & c,d & \cite{2022MNRAS.512.1522D} \\
NGC0524 & 23.3 & No & S0-a & - & 11.40 & z0MGS & -0.56 & e,g,h & \cite{2019MNRAS.485.4359S} \\
NGC0612 & 130.4 & Yes & S0-a & LINER & 11.76 & ${\rm M}_{K_s}$ & 0.85 & b,j & \cite{2023MNRAS.522.6170R} \\
NGC0708 & 58.3 & Yes & E & LERG & 11.75 & MASSIVE & -0.29 & e,h,i & \cite{2021MNRAS.503.5179N} \\
NGC1194 & 53.2 & No & S0-a & Seyfert 2 & 10.64 & z0MGS & -1.74 & j & This work \\
NGC1387 & 19.9 & No & E-S0 & LINER & 10.67 & z0MGS & -0.68 & d,e & Boyce, in prep \\
NGC1574 & 19.3 & No & E-S0 & - & 10.79 & z0MGS & -0.91 & c,e & \cite{2023MNRAS.522.6170R} \\
NGC2110 & 35.6 & No & E-S0 & - & 10.41 & ${\rm M}_{K_s}$ & $-1.41^2$ & c,k & This work \\
NGC3169 & 18.7 & Yes & Sa & Seyfert 1 & 10.84 & z0MGS & 0.29 & e,i & \cite{2022MNRAS.512.1522D} \\
NGC3351 & 10.0 & Yes & Sb & LERG & 10.28 & z0MGS & $-1.29^2$ & e,n & This work \\
NGC3368 & 18.0 & Yes & Sab & LERG & 10.67 & z0MGS & -0.29 & k & \cite{2022MNRAS.512.1522D} \\
NGC3607 & 22.2 & No & E-S0 & - & 11.34 & A3D & -0.54 & e,i & \cite{2022MNRAS.512.1522D} \\
NGC3862 & 92.5 & Yes & E & LERG & 11.68 & MASSIVE & $-0.63^2$ & a,e,i,l & This work \\
NGC4061 & 94.1 & Yes & E & - & 11.54 & MASSIVE & -0.71 & a,e,i,l & \cite{2022MNRAS.512.1522D} \\
NGC4261 & 31.9 & Yes & E & LINER & 10.80 & ${\rm M}_{K_s}$ & $-1.93^2$ & a,l & \cite{2023MNRAS.522.6170R} \\
NGC4429 & 16.5 & No & S0-a & - & 11.17 & A3D & -0.84 & e,n & \cite{2018MNRAS.473.3818D} \\
NGC4435 & 16.5 & No & S0 & - & 10.69 & A3D & -0.84 & e,i & \cite{2022MNRAS.512.1522D} \\
NGC4438 & 16.5 & No & Sa & LINER & 10.75 & z0MGS & -0.3 & e,i & \cite{2022MNRAS.512.1522D} \\
NGC4501 & 14.0 & Yes & Sb & Seyfert 2 & 11.00 & z0MGS & 0.43 & e,g & \cite{2022MNRAS.512.1522D} \\
NGC4697 & 11.4 & No & E & - & 11.07 & A3D & -1.08 & i & \cite{2017MNRAS.468.4675D} \\
NGC4826 & 7.4 & No & SABa & Seyfert 1 & 10.20 & z0MGS & -0.71 & e,n & \cite{2022MNRAS.512.1522D} \\
NGC5064 & 34.0 & No & Sb & - & 10.93 & z0MGS & 0.11 & e,g & \cite{2022MNRAS.512.1522D} \\
NGC5765b & 114.0 & No & SABb & Seyfert 2 & 11.21 & ${\rm M}_{K_s}$ & 1.43 & j & \cite{2022MNRAS.512.1522D} \\
NGC5806 & 21.4 & Yes & Sb & Seyfert 2 & 10.57 & z0MGS & -0.03 & d,e & \cite{2022MNRAS.512.1522D} \\
NGC5995 & 107.5 & No & SABa & Seyfert 2 & 11.41 & ${\rm M}_{K_s}$ & - & b & This work \\
NGC6753 & 42.0 & No & Sb & - & 10.78 & z0MGS & 0.32 & e,g & \cite{2022MNRAS.512.1522D} \\
NGC6958 & 30.6 & No & E & - & 10.76 & z0MGS & -0.58 & e,g & Thater, in prep \\
NGC7052 & 51.6 & Yes & E & $\rm LERG^1$ & 11.75 & MASSIVE & -0.07 & a,l & \cite{2021MNRAS.500.1933S} \\
NGC7172 & 33.9 & No & Sa & Seyfert 2 & 10.76 & z0MGS & 0.38 & m & \cite{2022MNRAS.512.1522D} \\
PGC043387 & 95.8 & No & E & - & 11.12 & ${\rm M}_{K_s}$ & -0.48 & i & This work \\
\hline
\end{tabular}
\parbox[t]{\textwidth}{\textit{Notes:} (1) galaxy name, (2) galaxy distance in Mpc, (3) whether a resolved radio jet is present in radio observations of the galaxy, (4) galaxy morphological type, (5) AGN type or HERG/LERG classification (determined using classification from Figure 2 of \citealt{2012MNRAS.421.1569B}). The radio AGN classification of NGC7052 was taken from \cite{2020ApJ...905...42G}. (6) galaxy stellar mass, (7) the reference for the stellar mass: L22 refers \cite{2022MNRAS.516.4066L}, C17 refers to \cite{2017GCN.21707....1C}, S10 refers to \cite{2010ApJ...714..713S}, A3D refers to \cite{2013MNRAS.432.1862C}, MASSIVE refers to \cite{2017MNRAS.471.1428V}, and z0MGS to \cite{2019ApJS..244...24L}. ${\rm M}_{K_s}$ refers to masses estimated from the galaxies $K_s$-band magnitude using Equation 2 of \cite{2013ApJ...778L...2C}. (8) the star formation rate of the galaxy. The uncertainties on the star formation rates are 0.2 dex for all sources except NGC0404, NGC1194 and PGC043387 which have uncertainties of 0.22, 0.87 and 0.22 dex respectively. $^2$ represents galaxies where the star formation rate was estimated in this work. (9) ALMA Project codes of each source, where a:\,2016.2.00046.S, b:\,2017.1.00904.S, c:\,2015.1.00419.S, d:\,2016.1.00437.S, e:\,2016.2.00053.S, f:\,2017.1.00572.S, g:\,2015.1.00466.S, h:\,2017.1.00391.S, i:\,2015.1.00598.S, j:\,2016.1.01553.S, k:\,2016.1.00839.S, l:\,2018.1.00397.S, m:\,2019.1.00363.S and n:\,2013.1.00493. (10) reference where the ALMA data were initially presented.}
\label{tab:Table 1}
\end{table*}

\subsection{ALMA observations and data reduction}
\label{Section 2.2}
Thirty-two sample objects were observed in $^{12}$CO(2-1) and 230~GHz continuum using ALMA Band 6, while three (NGC3351, NGC4429, NGC4826) have Band 7 $^{12}$CO(3-2) and 345~GHz continuum observations. 
The ALMA observations used in this work were taken between 2013 and 2020 as part of a large number of projects (see Table \ref{tab:Table 1}). For each target we used multiple ALMA observations with multiple array configurations. This enabled us to reach high angular resolution, while ensuring  adequate uv-plane coverage and excellent flux recovery. The spectral configuration always consisted of four spectral windows (SPWs), one centred on the redshifted frequency of the $\rm ^{12}CO$ line (rest frequency 230.5 GHz for the 2--1 transition, 345.8 GHz for 3--2). The other three SPWs were used to observe the continuum. ALMA data were reduced using the Common Astronomy Software Applications (CASA) pipeline \citep{2007ASPC..376..127M} version appropriate for each dataset. A standard calibration strategy was adopted for every observation. A single bright object (typically a quasar) was used as both flux and bandpass calibrator, while a second bright object was used as a phase calibrator. More details on the data reduction process can be found in \cite{2022MNRAS.512.1522D}.

\subsubsection{Line imaging}
\label{Section 2.2.2}
In this work we make use of the CO data cubes presented in \cite{2022MNRAS.512.1522D}, or used the same cleaning methods described therein for consistency. The final cleaned cubes have synthesised beam sizes ranging from 0$''$.054 to 0$''$.659, corresponding to spatial scales from 0.8 to 291~pc, and noise levels ranging from 0.19 $\rm mJy \ beam^{-1}$ to 3.70 $\rm mJy \ beam^{-1}$. Since we are interested only in the gas reservoirs on scales $\lesssim100$~pc (i.e.\,those relevant for the SMBH accretion process), we restrict our analysis to the sub-sample of 29 WISDOM galaxies whose ALMA data cubes satisfy such spatial resolution requirement.

\begin{figure*}
    \centering
    \includegraphics[width=0.9\textwidth]{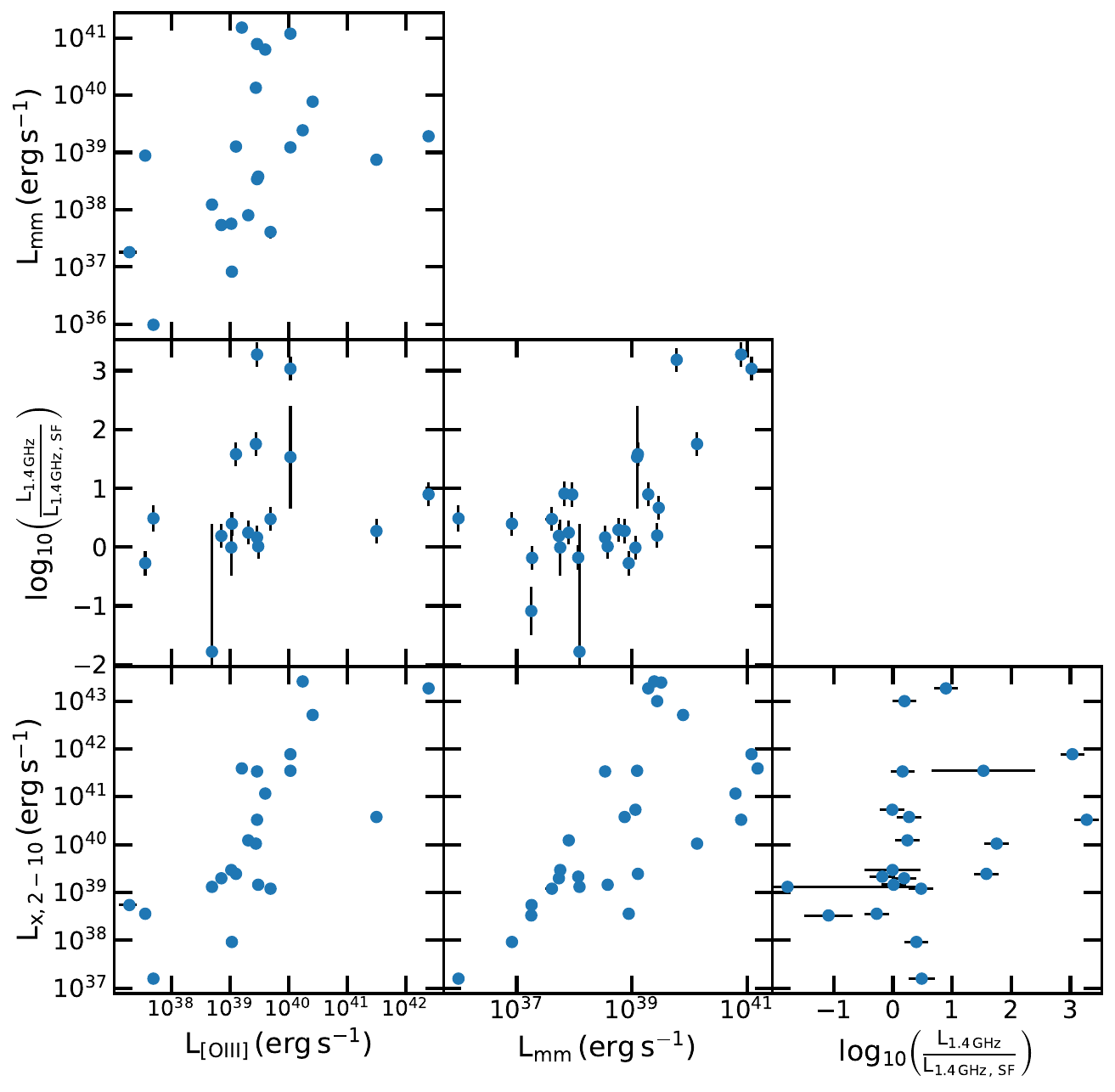}
    \caption{Correlations between the nuclear mm continuum luminosity, excess 1.4\,GHz factor, 2--10\,keV luminosity and the [OIII] luminosity. The Spearman rank coefficients and \textit{p}-value of these correlations are listed in Table \ref{tab:Table 3}.}
    \label{fig:Emission-correlation}
\end{figure*}

\subsection{Ancillary Data} 
\label{Section 2.3}
We gathered a variety of ancillary data to assess the level of nuclear activity in our targets and test its connection with the cold gas reservoirs observed with ALMA. For each source, we thus retrieved 2--10~keV X-ray, 1.4~GHz radio continuum and optical [OIII]$\lambda$5007 luminosities from the NASA Extragalactic Database (NED)\footnote{\url{https://ned.ipac.caltech.edu/}} or other literature sources. 

All the ancillary observations used in this work are listed in Table~\ref{tab:A1}. In the following, we briefly describe each set of ancillary data and caveats around their use. 

\par
\subsubsection{X-ray data}\label{sec:2.3.1}
We retrieved the nuclear 2--10~keV X-ray luminosities ($L_{X, 2-10})$ of the majority of our sources from \cite{2020ApJ...900..124B}, who presented a catalogue of nearby galaxies observed with \textit{Chandra}. Ten sample galaxies are not included in this catalogue\footnote{FRL49, FRL1146, NGC0449,NGC0524, NGC1194, NGC3351, NGC3368, NGC3862, NGC4429 and NGC4826}, thus their X-ray data was gathered from NED and comes from a variety of satellites, i.e.\,R\"{o}ntgensatellit (\textit{ROSAT}), \textit{EINSTEIN}, Advanced Satellite for Cosmology and Astrophysics (\textit{ASCA}), XMM-\textit{Newton} and \textit{Chandra} (see Table~\ref{tab:A1}). For four (NGC3351, NGC3862, NGC4429 and NGC4826) of these ten galaxies, 2--10 keV luminosities were not available, thus we scaled the available measurement (0.2--2 keV for NGC3351 and NGC4429, 0.3--8 keV for NGC3862 and NGC4826)  to the 2--10 keV energy band using a power law with an index $-0.8$ (corresponding to the mean reported by \citealt{2000MNRAS.316..234R}). Eight sample galaxies do not have any X-ray data available. 

{ Thanks to the exquisite {\it Chandra} resolution ($\approx0.5\arcsec$ on-axis) and the efforts by the authors to remove as much contamination as possible, the nuclear 2--10~keV luminosities from \cite{2020ApJ...900..124B} are expected to trace only emission from the unresolved AGN core (although some negligible contamination from unresolved nuclear X-ray sources may still occur)}. In the other ten cases, the spatial resolution of the available X-ray observations does not allow to distinguish between nuclear AGN emission and other types of contribution on larger galaxy scales, thus possible sources of contamination need to be considered. This includes emission from the diffuse hot atmospheres in and around galaxies (i.e.\,the circumgalactic medium, CGM). This low-surface brightness emission, however, usually requires very deep X-ray observations to be detected, and is typically dominant in the softer (0.3--2~keV) energy range. As such, we expect CGM contamination to be minimal even in low-resolution 2--10~keV X-ray data. A relatively larger contribution from the CGM may still be present in faint X-ray sources and in the four cases in which the 2--10 keV luminosities was extrapolated from lower energy bands.

Stellar X-ray binaries are another potential source of contamination. The K-band luminosity has been demonstrated to correlate with the luminosity of low mass X-ray binaries (LMXB). We thus use the $K_{s}$-band flux to estimate the contribution of LMXB to the 2--10~keV emission using the relations from \cite{2004ApJ...611..846K} and \cite{2011ApJ...729...12B}. We find that in three of the ten galaxies for which \textit{Chandra} data was not available from \cite{2020ApJ...900..124B} the contributions from LMXBs is minimal (<5\%). In the other seven galaxies (NGC0449, NGC0524, NGC3351, NGC3368, NGC3862, NGC4429 and NGC4826) the contribution expected from up to ~43\%. To estimate the potential contamination from high mass X-ray binaries (HMXB) we used the relation between SFR and X-ray luminosity from \cite{2003MNRAS.339..793G}. 
In this case, we find three galaxies have minimal contributions (<5\%), six (NGC0449, NGC0524, NGC3368, NGC3351, NGC4429 and NGC4826) have larger contributions up to ~73\%, and one (FRL1146) where we do not have information on the SFRs so we could not calculate the contributions from HMXB. We therefore assume that the 2-10~keV luminosity of most of our targets is dominated by AGN emission from the core. In few individual sources, however, it is possible that we slightly overestimate the AGN luminosity due to the aforementioned uncertainties. 


%

\subsubsection{Radio data}
\label{2.3.2}
We used 1.4~GHz radio observations to probe the type of nuclear activity, and the presence of radio jets in our sources. The data used in this work are mostly from the Very Large Array sky surveys, such as the Faint Images of the Radio Sky at Twenty-Centimeters (FIRST; \citealp{1994ASPC...61..165B}), and the NRAO VLA Sky Survey (NVSS; \citealp{1998AJ....115.1693C}). The spatial resolution of these surveys is typically very poor (e.g.\,45$\arcsec$ for the NVSS). While this does not allow us to resolve nuclear radio structures, it does ensures that no radio emission is resolved out and that any associated large-scale radio jet is detected. For the one source (NGC1574) for which 1.4 GHz observations were not available, we scaled the available 5GHz radio data to 1.4 GHz using a power law with a spectral index $\alpha=-0.8$ (for $S\propto\nu^{\alpha}$), as typical for optically-thin radio jet emission \citep[e.g.][]{1994A&A...285...27K,2013MNRAS.432.1114L}.

It is possible that some of the radio emission detected in our targets is contaminated by star formation within the galaxy (e.g.\,from supernova remnants). 
To quantify such putative contribution, we gathered the star formation rates (SFRs) of our sample sources from \cite{2022MNRAS.512.1522D}, when available, as before. For the 6 sample galaxies not included in that work, we estimated the SFRs adopting the following relation:
\begin{equation}\label{Eq 2}
    {\rm SFR}=\frac{M_{\rm H_2}}{\tau}
\end{equation}

where $M_{\rm H_2}$ is the total molecular gas mass withing the galaxy (calculated as described in Section~\ref{sec:3.1}), and $\tau$ is the depletion time, assumed to be 2~Gyr \citep[e.g][]{2008AJ....136.2782L}.
We then estimated the expected 1.4~GHz radio luminosity due to star formation using the following relation \citep{2011ApJ...737...67M}:
\begin{multline}
    \left(\frac{{\rm SFR}_{\nu}}{\rm M_{\odot}yr^{-1}}\right)= 10^{-27}\left[2.18\left(\frac{T_{e}}{\rm 10^4K}\right)^{0.45}\left(\frac{\nu}{\rm GHz}\right)^{-0.1}+15.1 \right.\\
    \left. \left(\frac{\nu}{\rm GHz} \right)^{\alpha^{\rm NT}} \right]^{-1}\left(\frac{L_{\nu}}{\rm erg \, s^{-1}Hz^{-1}}\right),
\end{multline}
which can be rearranged to:
\begin{multline}
    \left(\frac{L_{\nu}}{\rm erg \, s^{-1}Hz^{-1}}\right)=10^{27}\left(\frac{{\rm SFR}_{\nu}}{\rm M_{\odot}yr^{-1}}\right)\left[2.18\left(\frac{T_{e}}{\rm 10^4K}\right)^{0.45}\left(\frac{\nu}{\rm GHz}\right)^{-0.1}\right.\\+15.1 \left. \left(\frac{\nu}{\rm GHz} \right)^{\alpha^{\rm NT}}\right]
\end{multline}
where $\nu$ is the observed frequency, $T_e$ is the electron temperature and $\alpha^{\rm NT}$ is the non-thermal spectral index. We assumed $T_e=10^{4}$~K \citep{2011ApJ...737...67M} and $\alpha=-0.8$ \citep{2011ApJ...737...67M}. The relation combines thermal radio emission (calculated from the ionizing photon production rate) and non-thermal radio emission from supernovae, both of which are related directly to the SFR. 
\par
From the ratio between the total radio luminosity and that expected from star formation, ${\rm log_{10}}(L_{\rm 1.4GHz}/L_{\rm 1.4GHz, SF})$, we calculate what we call the radio excess factor ($E_{\rm 1.4}$). 
In galaxies with $E_{\rm 1.4}$ significantly larger than zero, the detected radio emission cannot be explained by star formation, and thus likely arises from nuclear activity. The radio excess factor for each galaxy is tabulated in Table \ref{tab:A1}. 

\subsubsection{Optical line data}\label{sec:2.3.3}
[O\,III]$\lambda$5007 is typically the strongest emission line in optical spectra of AGN and arises from gas in the narrow line regions (NLRs) that has been photo-ionised by the AGN radiation. It is then usually considered as a good proxy of the AGN bolometric luminosity \citep[e.g.][]{2014ARA&A..52..589H}. [O\,III] has also the advantage to be a more ubiquitous tracer of nuclear activity than the 2-10~keV luminosity, as it is observed in both kinetic- and radiative-mode AGN and does not suffer of any obscuration from the dusty torus (present in typical Seyfert-like objects). We therefore collected [OIII]$\lambda$5007 luminosities from a variety of instruments/surveys, such as the double spectrograph at the {\it Hale Telescope} \citep{1995ApJS...98..477H}, \textit{El Leoncito Astronomical Complex} \citep[CASLEO;][]{1997ApJ...486..132B,2000ApJS..126...63R}, the DOLORES
(Device Optimized for the LOw RESolution) spectrograph at \textit{Galileo National Telescope} \citep[TNG;][]{2009A&A...495.1033B}, \textit{Sloan Digital Sky Survey} \citep[SDSS;][]{2011ApJ...742...73Z}, \textit{MPG/ESO telescope} \citep{1993MNRAS.263..999T}, the spectrograph on the \textit{Shane Telescope} at Lick Observatory \citep{1986ApJ...301..727D,1999MNRAS.306..857C} and the \textit{CTIO Telescope} \citep{2010ApJS..190..233M}.
We note that the [OIII]$\lambda$5007 line may be contaminated by star formation or old stars, or be affected by extinction arising within the host galaxy. However, star formation is only expected to  contribute significantly in higher-redshift galaxies (whereas it should be negligible in nearby objects such as our sample sources; \citealp{2016MNRAS.462..181S}). {Where available, we additionally gathered H$\beta$, [N\,II]$\lambda$6583 and H$\alpha$ luminosities. These are useful to calculate the [OIII]/H$\beta$ and [NII]/H$\alpha$ ratios which we can use to construct the BPT diagrams of our objects, and thus asses their dominant excitation mechanism \citep[e.g.][]{1981PASP...93....5B,Kewley06}. Eighteen of our sample galaxies have all the lines required to construct a BPT diagram, showing that 5 sources fall in the AGN-dominated region (FRL49, NGC0612,NGC1194,NGC2110, NGC5765b), 4 in the LINER region (NGC3368, NGC3862, NGC5995, NGC6753), 8 in the composite region (NGC1387, NGC3351, NGC4061, NGC4261, NGC4826, NGC5064, NGC7172, PGC043387) and 1 (MRK567) in the SF-dominated region. The log([OIII]/H$\beta$) ratios for 9 other galaxies (FRL1146, NGC0404, NGC0524, NGC3169, NGC3607, NGC4429, NGC4435, NGC4438, NGC4501) have values ranging from -0.44 to 0.71. Depending on their unknown log([NII]/H$\alpha$) ratios, they could thus be placed in the star formation, composite, LINER regions or AGN regions.}

\subsubsection{ALMA nuclear continuum emission}
\label{3.4}
We gathered the ALMA nuclear continuum luminosities from \citet{Ruffa23b}, with the mm continuum flux which have been measured from the innermost beam at the position of the AGN in the ALMA continuum map of each galaxy. 



\subsubsection{Accretion tracer correlations}
The correlations between the nuclear mm luminosity, X-ray luminosity, [OIII] luminosity and excess radio factor are shown in Figure \ref{fig:Emission-correlation}, with the correlation coefficients and p-values listed in Table \ref{tab:Table 3}. This figure shows that the four tracers of activity mostly correlate with each other despite different contaminants, suggesting we are tracing nuclear activity rather then larger scale emission. This also shows that, even though some of the galaxies in our sample are not formally classified as AGN, low-level nuclear activity seems to be present. 


\subsubsection{Stellar masses}
The stellar masses of the majority of our sample galaxies were taken from \cite{2022MNRAS.512.1522D}, who in turn collected them from the ATLAS\textsuperscript{3D} \citep{2013MNRAS.432.1862C} and MASSIVE \citep{2017MNRAS.471.1428V} surveys, and the \textit{z}=0 Multiwavelength Galaxy Synthesis (z0MGS) project \citep{2019ApJS..244...24L}. The stellar masses for MRK\,567 and NGC\,0404 were taken from \cite{2017GCN.21707....1C} and \cite{2010ApJ...714..713S}, respectively. Where stellar mass measurements were not available in the literature, we estimated them from the $K_s$-band magnitudes measured in the extended source catalogue of the 2 micron All-Sky Survey \citep[2MASS;][]{2003AJ....125..525J}. We used Equation 2 of \cite{2013ApJ...778L...2C}, with no correction for the emission from the AGN (as this should be small at these frequencies in our low-luminosity sources):
\begin{equation}
    {{\rm log}_{10}}M_* \approx 10.58-0.44 \times (M_{K_s}+23)
\end{equation}
where $M_*$ is the stellar mass and $M_{K_s}$ is the $K_s$-band magnitude. 

\section{Methodology and derived quantities}
\label{Section 3}

In this work we search for correlations between \textit{circumnuclear} molecular gas reservoirs and SMBH fuelling across a sample of galaxies with a range of nuclear activities. This requires us to assess both the amount of molecular gas present in the circumnuclear regions, and its structure. Furthermore, we need to constrain the SMBH accretion rate in our sources, and the type of nuclear activity. Below we describe the methodology we adopted to determine these quantities. 

\subsection{Molecular gas masses}
\label{sec:3.1}
We adopt the following relation to estimate the molecular hydrogen gas masses ($M_{\rm H_2}$) of our galaxies within different apertures \citep{2013ARA&A..51..207B}:
\begin{equation}
    M_{\rm H_2}=2{\rm m_H}\frac{\lambda^2}{2k_{\rm B}}X_{\rm CO}D_{\rm L}^2R\int S_\nu\,dV,
    \label{eq:mol_mass1}
\end{equation}
where $\rm m_{\rm H}$ is the mass of the hydrogen atom, $\lambda$ is the rest wavelength of the observed molecular transition, $k_{\rm B}$ is the Boltzmann constant, $X_{\rm CO}$ is the CO-to-$\rm H_{\rm 2}$ conversion factor, $D_{\rm L}$ is the luminosity distance, $R \equiv T_{\rm b,ref}/T_{\rm b,CO(1-0)}$ is the line intensity ratio (i.e. the ratio between the ground state and the observed CO line brightness temperature), and $\int S_{\rm v} d \rm V$ is the integrated flux density of the CO(1-0) line, with units matching those of $X_{\rm CO}$. This was estimated by integrating the spectrum of the observed CO transition within a given aperture over all the velocity channels of the line. Equation~\ref{eq:mol_mass1} can be simplified to
\begin{equation}
    \left(\frac{M_{\rm H_{2}}}{\rm M_{\odot}}\right)=7847 \, J_{\rm upper}^{-2} X_{\rm CO, 2\times10^{20}} R \left(\frac{D_{\rm L}}{\rm Mpc}\right)^2 \left(\frac{\int S_{\rm v} d \rm V}{\rm Jy \ km s^{-1}}\right), 
\end{equation}
where $J_{\rm upper}$ is the upper state rotational quantum number of the observed transition (here $J_{\rm upper}$ is 2 or 3) and $X_{\rm CO, 2\times10^{20}}$=$\frac{X_{\rm CO}}{\rm 2\times10^{20} \, {\rm cm^{-2}}(K \ km  \ s^{-1})^{-1}}$. As most of our galaxies are massive and metal-rich, we assume a Milky Way-like CO-to-$\rm H_{\rm 2}$ conversion factor of $\num{3e20}{\rm cm^{-2}}(\rm K \ km \ s^{-1})^{-1}$ \citep{1988A&A...207....1S}. We also assume the line ratios to be $T_{\rm b, CO(2-1)}/T_{\rm b, CO(1-0)}=0.7$ and $T_{\rm b, CO(3-2)}/T_{\rm b, CO(1-0)}=0.3$ \citep[see e.g.][]{2022ApJ...927..149L}. 
For sample galaxies observed with ALMA at adequate spatial resolutions, we estimated the molecular gas mass within three different elliptical apertures of radii 100, 75, and 50~pc. We additionally measure the molecular gas mass in a elliptical aperture with a radius of 200~pc to measure the nuclear molecular gas concentration. 
The resulting M$_{\rm H_{\rm 2}}$ are listed in Table \ref{tab:Table 2}. As mentioned above, the molecular gas masses were all calculated with a fixed X$_{\rm CO}$ factor. We note, however, that the CO-to-$\rm H_{2}$ conversion factors in galaxy centres may vary \citep{2013ApJ...777....5S}, adding uncertainty at a $\sim$0.3 dex level. To aid comparisons with different \textit{X}$_{\rm CO}$ prescriptions, we provide the CO integrated flux densities measured within the different apertures in Table~\ref{tab:A3}. We assume an additional 10\% error on our molecular gas masses due to ALMA calibration uncertainties. 

To assess the structure of the molecular gas at the centre of each sample galaxy, following \cite{2021A&A...652A..98G}, we also calculated the molecular concentration parameter:

\begin{equation}
\frac{\Sigma^{\rm 50pc}_{\rm H_2}}{\Sigma^{\rm 200pc}_{\rm H_2}} = 16\left(\frac{M^{\rm 50pc}_{\rm H_2}}{M^{\rm 200pc}_{\rm H_2}}\right),
\end{equation}
where $\Sigma^{x\mathrm{pc}}_{\rm H_2}$ and $M^{x\mathrm{pc}}_{\rm H_2}$ are the molecular gas surface density and mass, respectively, within a elliptical aperture of radius $x$~pc. We assumed that the gas lies within a flat disc so $\Sigma^{x\mathrm{pc}}_{\rm H_2}$=$\frac{M^{x\mathrm{pc}}_{\rm H_2}}{\pi x^2}$.

\begin{table*}
\caption{Circumnuclear masses for different region sizes and millimetre spectral indices for our sources.}
\label{tab:Table 2}
\sisetup{input-symbols = {( )}, parse-numbers=false}
\begin{tabular}{lllllllllllllll 
r
l}
\hline
{Galaxy} & {\thead{${\rm log}\left(\frac{ M_{\rm H_2}}{\rm M_\odot}\right)$}} & {\thead{$\Delta$ log M$_{\rm H_2}$}} & {\thead{${\rm log}\left(\frac{ M_{\rm H_2}}{\rm M_\odot}\right)$}} & {\thead{$\Delta$ log M$_{\rm H_2}$}} & {\thead{${\rm log}\left(\frac{ M_{\rm H_2}}{\rm M_\odot}\right)$}} & {\thead{$\Delta$ log M$_{\rm H_2}$}} & {\thead{${\rm log}\left(\frac{ M_{\rm H_2}}{\rm M_\odot}\right)$}} & {\thead{$\Delta$ log M$_{\rm H_2}$}} & {\thead{$\rm S_{\nu, mm}$}} & {\thead{$\rm \sigma S_{\nu,mm}$}} 
\\
& \multicolumn{2}{c}{(200\,pc)} & \multicolumn{2}{c}{(100\,pc)} & \multicolumn{2}{c}{(75\,pc)} & \multicolumn{2}{c}{(50\,pc)} & & 
\\ 
& & {(dex)} & & {(dex)} & & {(dex)} & & {(dex)} & {(mJy)} & {(mJy)} & &  \\
{(1)} & {(2)} & {(3)} & {(4)} & {(5)} & {(6)} & {(7)} & {(8)} & {(9)} & {(10)} &{(11)} 
\\
\hline
FRL49 & 8.34 & 0.04 & 7.87 & 0.04 & - & - & - & - & 0.93 & 0.038 
\\
FRL1146 & - & - & - & - & - & - & - & - & 0.47 & 0.0272 
\\
MRK567 & 8.90 & 0.04 & 8.52 & 0.04 & - & - & - & - & <0.11 & 0.0354 
\\
NGC0383 & 8.15 & 0.04 & 7.71 & 0.04 & 7.48 & 0.04 & 7.08 & 0.04 & 63 & 0.101 
\\
NGC0404 & 6.18 & 0.04 & 6.18 & 0.04 & 6.17 & 0.04 & 6.14 & 0.04 & 0.38 & 0.012 
\\
NGC0449 & - & - & - & - & - & - & - & - & 0.60 & 0.0246 
\\
NGC0524 & 7.42 & 0.04 & 6.93 & 0.05 & 6.75 & 0.04 & 6.47 & 0.04 & 5.7 & 0.023 
\\
NGC0612 & 8.20 & 0.04 & 7.62 & 0.04 & 7.50 & 0.04 & 7.35 & 0.05 & 25 & 0.06 
\\
NGC0708 & 8.12 & 0.04 & 7.66 & 0.04 & 7.47 & 0.04 & 7.13 & 0.05 & 1.3 & 0.0165 
\\
NGC1194 & 7.48 & 0.04 & 7.13 & 0.04 & 6.95 & 0.05 & - & - & 1.6 & 0.0286 
\\
NGC1387 & 7.66 & 0.04 & 7.05 & 0.04 & 6.80 & 0.04 & 6.43 & 0.04 & 1.0 & 0.0535 
\\
NGC1574 & 6.79 & 0.04 & 6.74 & 0.04 & 6.66 & 0.04 & 6.40 & 0.04 & 3.3 & 0.033 
\\
NGC2110 & 7.35 & 0.04 & 6.77 & 0.04 & - & - & - & - & 21 & 0.453 
\\
NGC3169 & 8.26 & 0.04 & 7.80 & 0.04 & 7.60 & 0.04 & - & - & 3.4 & 0.107 
\\
NGC3351 & 7.66 & 0.04 & 7.39 & 0.04 & 7.16 & 0.04 & 6.91 & 0.04 & <0.45 & 0.148 
\\
NGC3368 & 8.36 & 0.04 & 7.87 & 0.04 & 7.68 & 0.05 & 7.42 & 0.04 & <0.56 & 0.202 
\\
NGC3607 & 7.89 & 0.04 & 7.51 & 0.04 & 7.33 & 0.04 & - & - & 2.7 & 0.164 
\\
NGC3862 & - & - & - & - & - & - & - & - & 64 & 1.43 
\\
NGC4061 & 7.65 & 0.04 & 7.10 & 0.04 & 6.90 & 0.04 & - & - & 2.4 & 0.18 
\\
NGC4261 & 7.32 & 0.04 & 7.29 & 0.04 & 7.21 & 0.04 & 7.02 & 0.04 & 220 & 1.53 
\\
NGC4429 & 7.31 & 0.04 & 6.64 & 0.04 & 6.38 & 0.04 & 5.88 & 0.04 & 1.1 & 0.0853 
\\
NGC4435 & 7.61 & 0.04 & 7.25 & 0.04 & 7.05 & 0.05 & 6.78 & 0.04 & 0.73 & 0.0246 
\\
NGC4438 & 8.19 & 0.04 & 7.78 & 0.04 & 7.58 & 0.04 & 7.24 & 0.04 & 0.52 & 0.126 
\\
NGC4501 & 8.06 & 0.04 & 7.72 & 0.04 & 7.52 & 0.05 & 7.22 & 0.04 & 1.4 & 0.0789 
\\
NGC4697 & 6.02 & 0.04 & 5.93 & 0.04 & 5.83 & 0.04 & 5.61 & 0.04 & 0.48 & 0.0444 
\\
NGC4826 & 7.66 & 0.04 & 7.55 & 0.04 & 7.25 & 0.04 & 7.60 & 0.04 & 0.38 & 0.0746 
\\
NGC5064 & 8.13 & 0.04 & 7.67 & 0.04 & 7.41 & 0.04 & 7.10 & 0.04 & 0.28 & 0.0259 
\\
NGC5765b & 8.41 & 0.04 & - & - & - & - & - & - & 0.328 & 0.0616 
\\
NGC5806 & 7.57 & 0.04 & 7.16 & 0.04 & 7.02 & 0.04 & 6.77 & 0.05 & <0.14 & 0.0473 
\\
NGC5995 & - & - & - & - & - & - & - & - & 0.99 & 0.0331 
\\
NGC6753 & 8.63 & 0.04 & 8.16 & 0.05 & 7.94 & 0.04 & 7.62 & 0.04 & <0.14 & 0.0447 
\\
NGC6958 & 7.70 & 0.04 & 7.17 & 0.04 & 6.96 & 0.04 & 6.61 & 0.04 & 11 & 0.0569 
\\
NGC7052 & 7.75 & 0.04 & 7.39 & 0.04 & 7.17 & 0.04 & 6.79 & 0.04 & 18 & 0.0823 
\\
NGC7172 & 8.36 & 0.04 & 7.48 & 0.04 & 7.14 & 0.04 & 6.74 & 0.04 & 8.4 & 0.32 
\\
PGC043387 & - & - & - & - & - & - & - & - & <0.31 & 0.104 
\\ 
\hline
\end{tabular}
\parbox[t]{0.78\textwidth}{\textit{Notes:} (1) galaxy name. (2) mass measured within a elliptical aperture of 200\,pc radius, with its uncertainty in (3). (4)-(9) follow the same pattern, for 100, 75 and 50\,pc apertures. (10) nuclear mm continuum flux derived from all our ALMA data (fluxes measured separately from the lower and upper sidebands in Table \ref{Appendix A}), (11) nuclear mm continuum flux uncertainty. 
}
\end{table*}

\subsection{Accretion rates}
\label{3.2}
To estimate the SMBH accretion rates in each source, we use the following relation \citep{2012NewAR..56...93A}: 
\begin{equation}
    \left(\frac{\dot{M}_{\rm BH}}{\rm M_{\odot \, }yr^{-1}}\right)=0.15\left(\frac{0.1}{\eta}\right) \left(\frac{ L_{\rm Bol}}{\rm 10^{45} \, ergs \ s^{-1}}\right)
\end{equation}
where $\rm L_{Bol}$ is the AGN bolometric luminosity and $\eta$ is the mass-energy conversion efficiency factor, typically assumed to be 0.1 \citep[e.g.][]{2004MNRAS.351..169M}. 

We follow two prescriptions to estimate the AGN bolometric luminosities.
The 2--10~keV X-ray luminosity is generally considered a good proxy of the AGN bolometric luminosity in radiative-mode AGN \citep[e.g.][]{2008ARA&A..46..475H}, as X-ray emission in these sources is expected to come from the corona above the accretion disc. In this case, the bolometric luminosity can be calculated by inverting the bolometric correction relation of \citet{2004MNRAS.351..169M}:
\begin{equation}
    {\rm log}_{10}\left(\frac{L_{\rm Bol}}{L_{2-10keV}}\right)=1.54+0.24\mathcal{L}+0.012\mathcal{L}^2-0.0015\mathcal{L}^3,
\end{equation} 
where $\mathcal{L} \equiv {\rm log}_{10}\left(\frac{L_{\rm Bol}}{ L_{\odot}}\right)-12$. 

In kinetic-mode AGN, the accretion disc is expected to be absent and the 2--10~keV emission may instead arise from inverse Compton up-scattering of non-thermal photons from the radio jet \citep[e.g.][]{2006ApJ...644L..13B}. {For this reason, we additionally used the [OIII]$\lambda$5007 luminosity as a proxy of $\rm L_{Bol}$, adopting the bolometric correction $\rm L_{bol}/L_{[OIII]}\approx3500$ \citep{2004ApJ...613..109H}, to ensure both types of AGN are covered.} [OIII]$\lambda$5007 is a more ubiquitous tracer of nuclear activity as it is usually the brightest emission-line in optical spectra of AGN and is less contaminated than other emission lines. As illustrated in Table~\ref{tab:Table 1}, a clear AGN classification is missing for some of our sample galaxies (13/35), whereas the majority of them consist on a mix of radiative- and kinetic-mode AGN. For each source, we thus calculate $L_{\rm bol}$ using both the 2--10~keV and [OIII]$\lambda$5007 luminosity as tracers. We show the relation between these two derived bolometric luminosities in Figure \ref{fig:X-ray-OIII}. We note there is reasonably large scatter in the two measures of bolometric luminosity. This is likely due to the mix of radiative- and kinetic-mode objects in our sample, with the X-ray emission not being a good proxy for the bolometric luminosity of the latter (see above). 

\begin{table*}
    \centering
    \caption{Spearman rank coefficients and $p$-values}
    \begin{tabular}{cc
    S[table-format=1.4]
    S[table-format=1.2]
    S[table-format=1.2]
    S[table-format=1.2]
    S[table-format=1.4]}
         \hline
         & \multicolumn{6}{c}{Mass--Luminosity Correlations} \\
         \hline
         & \multicolumn{2}{c}{log ($M_{\rm H_2, 100pc}/\rm M_\odot)$} & \multicolumn{2}{c}{log ($M_{\rm H_2, 75pc}/\rm M_\odot$)} & \multicolumn{2}{c}{log ($M_{\rm H_2, 50pc}/\rm M_\odot$)}  \\
         & \multicolumn{2}{c}{(1)} & \multicolumn{2}{c}{(2)} & \multicolumn{2}{c}{(3)} \\
         \hline
         & {Coefficient} & {\textit{p}-value} & {Coefficient} & {\textit{p}-value} & {Coefficient} & {\textit{p}-value} \\
         \hline
         log($E_{\rm 1.4}$) & 0.32 & 0.12 & 0.35 & 0.11 & 0.56 & 0.01 \\
         log($L_{\rm X,2-10}$) & 0.31 & 0.16 & 0.29 & 0.21 & 0.29 & 0.26 \\
         log($L_{\rm mm}$) & 0.002 & 0.99 & 0.14 & 0.55 & 0.11 & 0.66 \\
         log($\dot{M}_{\rm acc,X-ray}$) & 0.31 & 0.16 & 0.29 & 0.21 & 0.29 & 0.26 \\
         log($\dot{M}_{\rm acc,[OIII]}$) & 0.27 & 0.26 & 0.38 & 0.13 & 0.53 & 0.05 \\
         \hline
         & \multicolumn{6}{c}{Luminosity--Luminosity Correlations}\\
         \hline
         & \multicolumn{3}{c}{Coefficient} & \multicolumn{3}{c}{\textit{p}-value}\\
         & \multicolumn{6}{c}{(4)} \\
         \hline
         $L_{\rm mm}-L_{\rm X,2-10}$ & \multicolumn{3}{c}{0.76} & \multicolumn{3}{c}{3.37E-6} \\
         $L_{\rm mm}-E_{\rm 1.4}$ & \multicolumn{3}{c}{0.57} & \multicolumn{3}{c}{0.008} \\
         $E_{\rm 1.4}-L_{\rm X, 2-10}$ & \multicolumn{3}{c}{0.16} & \multicolumn{3}{c}{0.56}\\
         $L_{\rm [OIII]}-L_{\rm X,2-10}$ & \multicolumn{3}{c}{0.78} & \multicolumn{3}{c}{2.27E-5}\\
         $L_{\rm [OIII]}-E_{\rm 1.4}$ & \multicolumn{3}{c}{0.15} & \multicolumn{3}{c}{0.60}\\
         $L_{\rm [OIII]}-L_{\rm mm}$ & \multicolumn{3}{c}{0.52} & \multicolumn{3}{c}{0.01}\\
         \hline
    \end{tabular}
     \parbox[t]{0.68\textwidth}{\textit{Notes:} (1) Spearman rank correlation coefficients and $p$-value for the 100~pc radius aperture, (2) and (3) same quantities for the 75 and 50~pc radius apertures. (4) lists the Spearman rank correlation coefficients and $p$-values between the luminosities studied.}
\label{tab:Table 3}
\end{table*}

\label{sec:results} 

\section{Results and Analysis} 
\label{Section 4}

\subsection{Properties of WISDOM AGN}
\label{samplefigs}

\begin{figure*}
    \centering
    \includegraphics[width=0.45\textwidth]{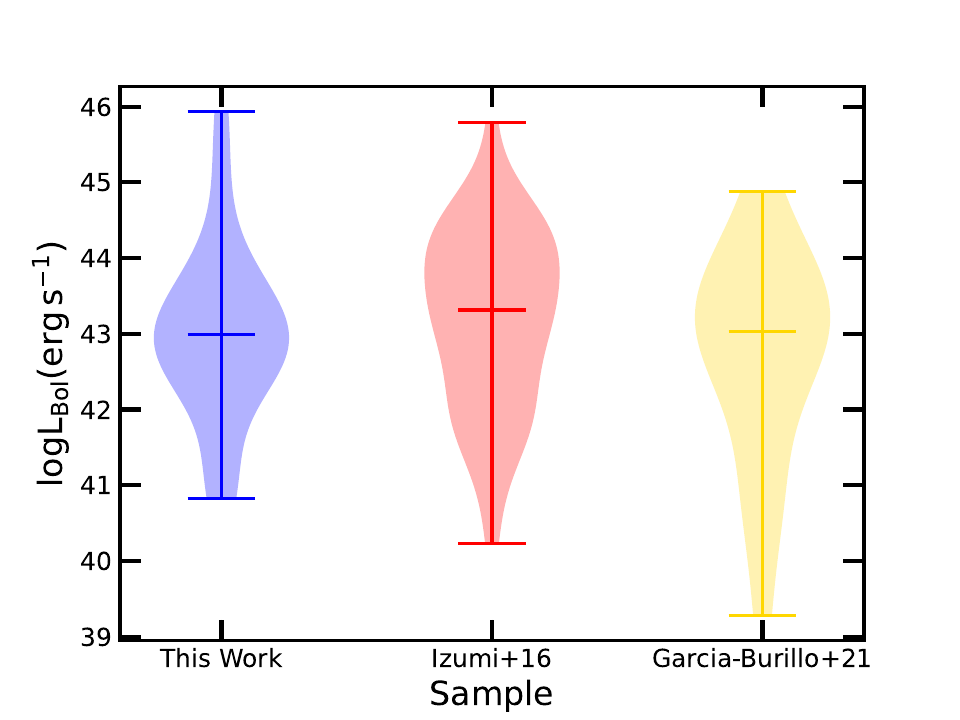}
    \includegraphics[width=0.45\textwidth]{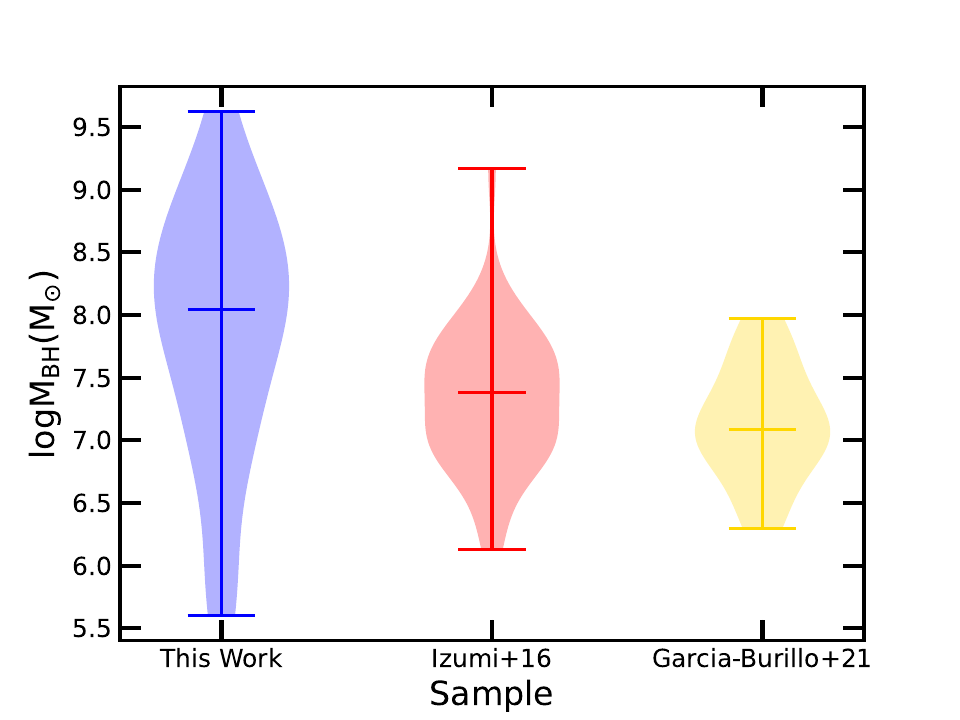}
    \includegraphics[width=0.45\textwidth]{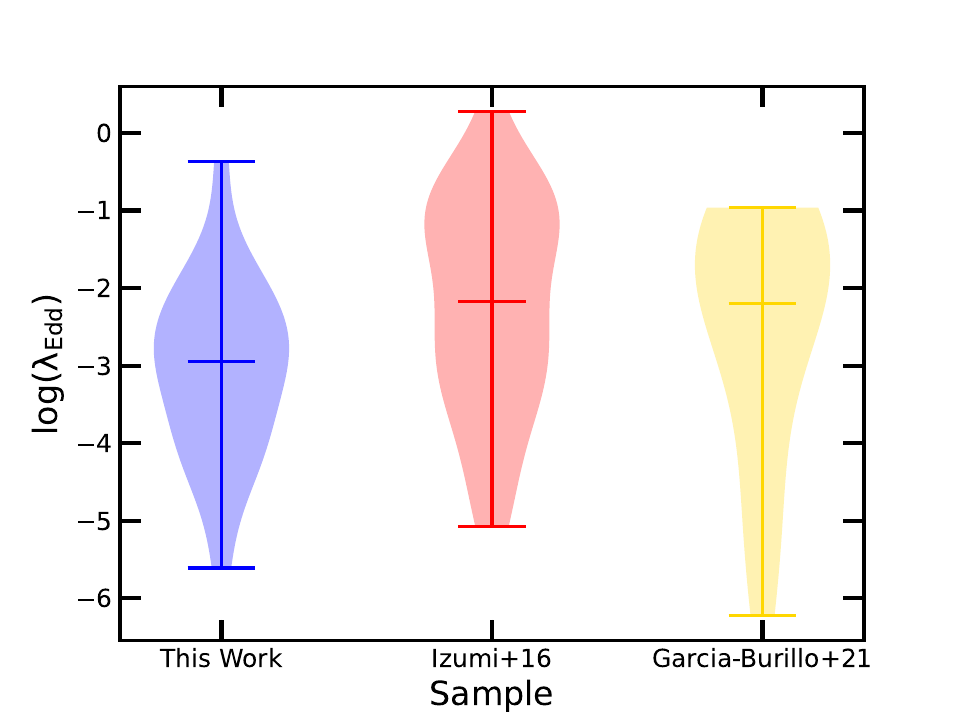}
    \includegraphics[width=0.45\textwidth]{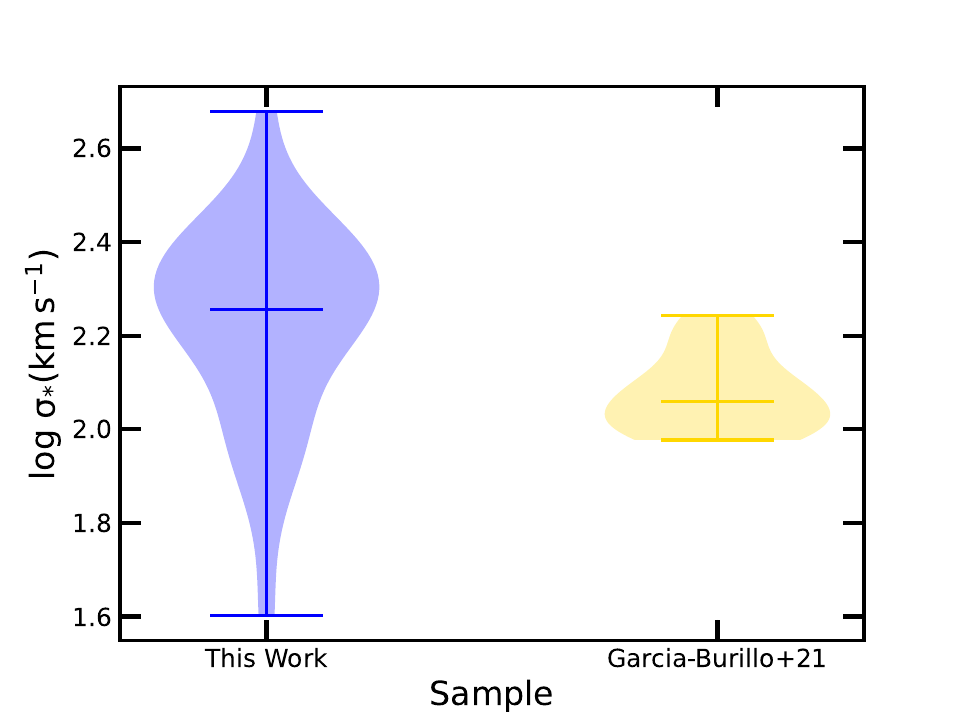}
    \includegraphics[width=0.45\textwidth]{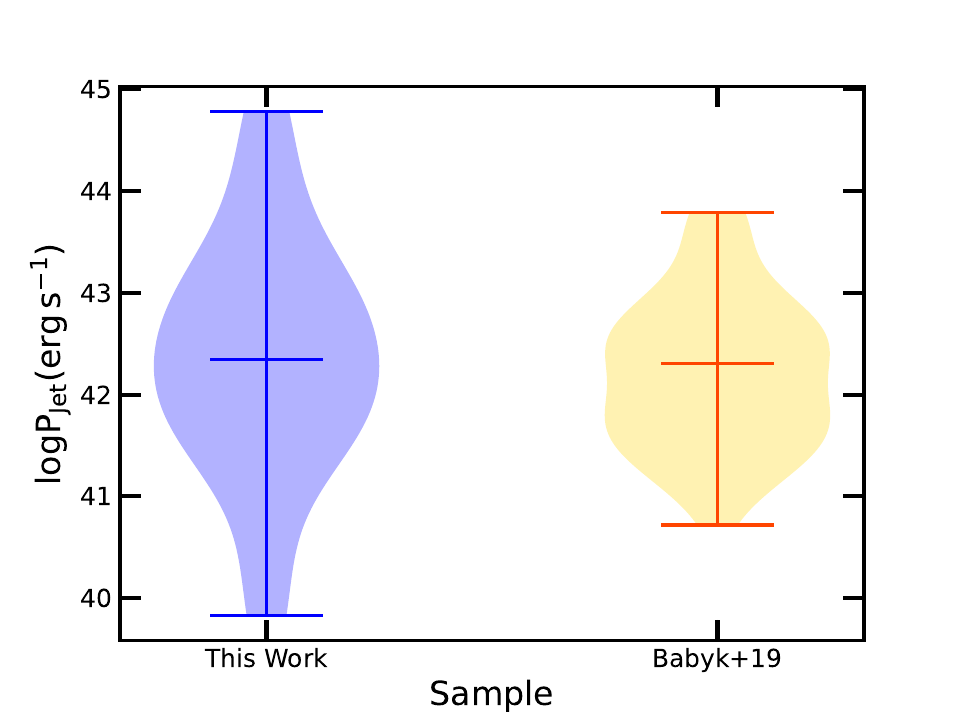}
    \caption{Distribution of AGN bolometric luminosity (top left), black hole mass $M_{BH}$ (top right), Eddington ratio $\lambda_{\rm Edd}$ (middle left), velocity dispersion (middle right),  and jet power (bottom) of the WISDOM sample. These are compared with the properties of the AGN from the works of  \protect \cite{Izumi16}, \protect \cite{Babyk19} and \protect \cite{2021A&A...652A..98G} which are discussed further in the text. The horizontal lines represent the median of each distribution.}
    \label{fig:Hists}
\end{figure*}

As discussed above, in this work we aim to investigate previous claims that at circumnuclear scales (<100\,pc) accretion rate tracers correlate with the mass and structure of the cold molecular gas mass.  
As illustrated in Table~\ref{tab:Table 1}, our sample consists on systems with a diverse range of nuclear activities. In order to further place the WISDOM galaxies in context with previous studies, in Figure \ref{fig:Hists} we illustrate the main properties of the AGN in our sample (i.e.\,bolometric luminosity, black hole mass, Eddington ratio \footnote{The Eddington ratio is a measure of the level of nuclear activity and is defined as $\lambda_{\rm Edd}=L_{\rm bol}/L_{\rm Edd}$, where $L_{\rm bol}$ is the AGN bolometric luminosity and $L_{\rm Edd}=1.26 \times 10^{38} M_{\rm BH}$~erg~s$^{-1}$ is the Eddington luminosity.} , central velocity dispersion and jet power), compared with those from the works of \protect \cite{Izumi16}, \protect \cite{Babyk19} and \protect \cite{2021A&A...652A..98G}, with the medians for the samples used and the KS test \textit{p}-values between the sample are shown in Table \ref{tab:stats}. In this work we calculated the jet power using the same method as \cite{Babyk19}. We calculated the radio power using the relation:
\begin{equation}
    P_{\nu_0}=4 \pi D_{\rm L}^{2}(1+z)^{\alpha-1}S_{\nu_0}\nu_0.
\end{equation}
This was then used to calculate the jet power using this relation from \cite{2010ApJ...720.1066C}:
\begin{equation}
    {\rm log}\,P_{\rm cav}=0.75\,{\rm log}\,P_{1.4}+1.91.
\end{equation}
It is clear from Figure~\ref{fig:Hists} that the bolometric luminosities of our AGN are consistent with those of the samples studied by \cite{Izumi16} and \cite{2021A&A...652A..98G}, and we probe a range of radio jet powers similar to that of the sources analysed by \cite{Babyk19}. On the other hand, the SMBH masses of the AGN in our sample  are larger - on average - than those probed in such previous studies, and thus their Eddington ratios are slightly lower (at least when compared with the work of \citealt{Izumi16}). 

More generally, Figure~\ref{fig:Hists} shows that there is overlap between the main properties of the AGN in our sample and those in the previous reference studies, with the WISDOM objects being also clearly complementary to such works. In the following, we will further discuss potential differences and if/how these may affect our results.

\subsection{AGN luminosity -- molecular gas mass correlations} 
\label{Section 4.1}

In Figures~\ref{fig:Figure 1}-\ref{fig:Figure 3} we show the obtained circumnuclear $\rm H_2$ masses plotted against excess 1.4~GHz continuum, 2-10~keV X-ray and nuclear mm-continuum luminosity, respectively. 
To check for the statistical significance of such relations, we carried out a Spearman rank analysis, where we consider relations with \textit{p}-values$\lesssim0.05$ as statistically significant. The resulting Spearman rank coefficients and \textit{p}-values are presented in Table \ref{tab:Table 3}
\subsubsection{Excess radio luminosity-- molecular mass correlation} 
\label{4.1.1}
\begin{figure*}
    \centering
    \includegraphics[width=18cm]{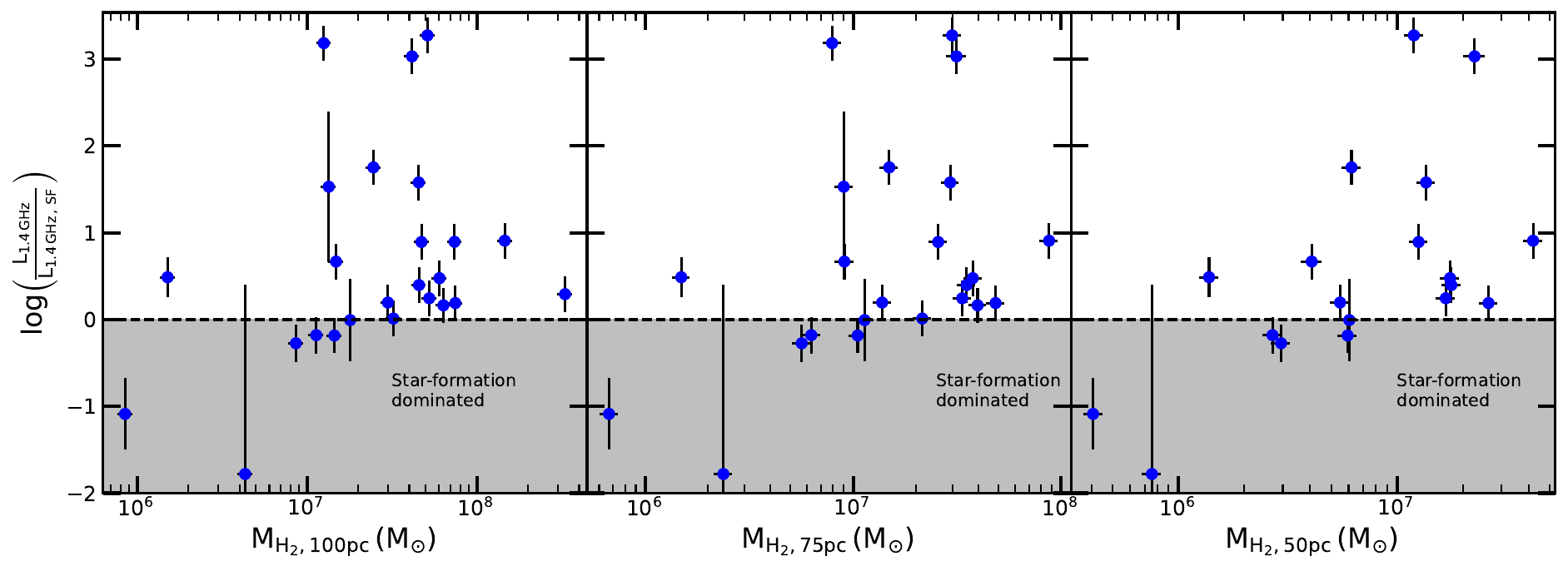}
    \caption{Molecular gas mass within an aperture of a given radius (100, 75 and 50~pc, as indicated by the x-axis labels) versus excess 1.4 GHz radio continuum fraction (after the contribution from star formation has been removed). 
    }
    \label{fig:Figure 1}
\end{figure*}

We show in Figure~\ref{fig:Figure 1} the correlation between molecular gas mass on sub-kpc scales and excess radio emission. 
There is no strong correlation between these quantities, as indicated by Spearman rank analysis (reported in the first row of Table \ref{tab:Table 3}). The correlation coefficient is 0.32 for the 100~pc radius aperture, and increases to 0.35 for the 75~pc and 0.56 for the 50~pc radius aperture, with \textit{p}-values of 0.12, 0.11 and 0.01, respectively. We note that, based on these results, a mild correlation may be present at the 50~pc scale. However, we checked that this is driven by galaxies that are dominated by star formation. When these objects are excluded, the Spearman rank coefficient becomes -0.07, with a \textit{p}-value of 0.82, thus finding no evidence for any correlation.

\subsubsection{X-ray luminosity-molecular gas mass correlation} 
\label{4.1.2}
\begin{figure*}
    \centering
    \includegraphics[width=18cm]{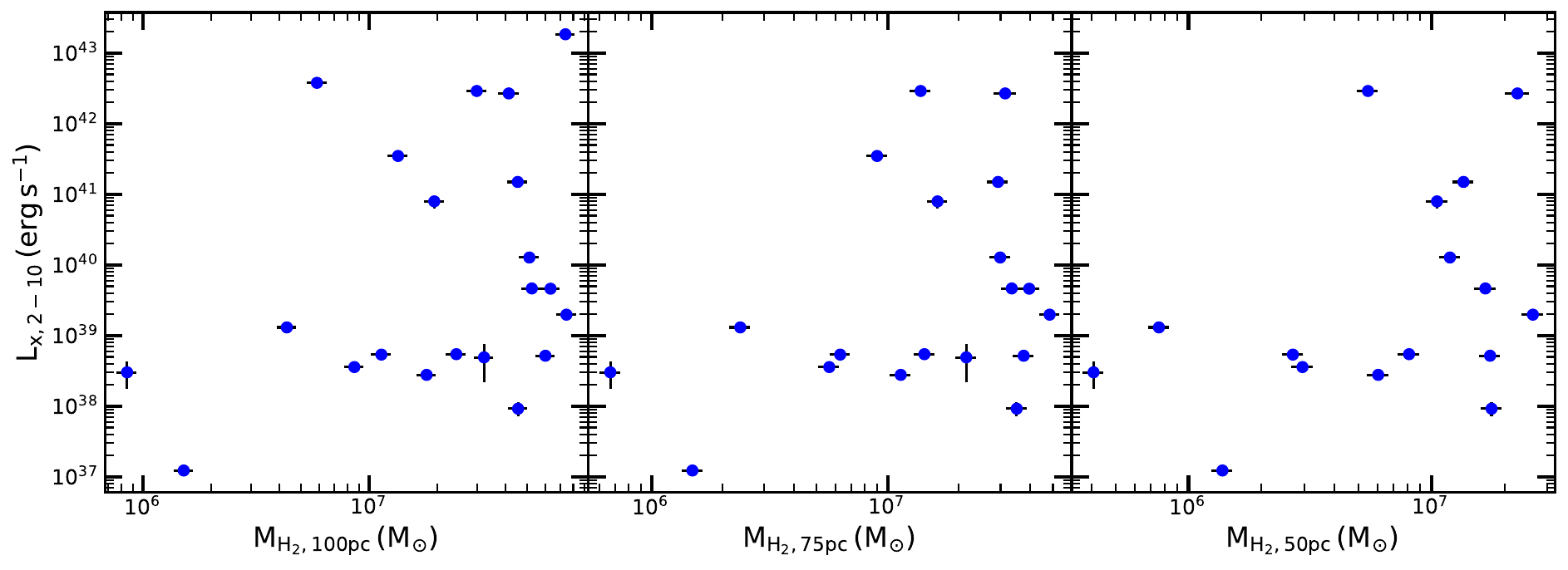}
    \caption{As Figure~\ref{fig:Figure 1}, but for the 2--10 keV X-ray luminosity. 
    }
    \label{fig:Figure 2}
\end{figure*}

We show in Figure~\ref{fig:Figure 2} the relation between the molecular gas mass on sub-kiloparsec scales and 2-10~keV X-ray luminosity. Also in this case, there is no significant correlation between the two quantities within any aperture size, as supported by the Spearman rank analysis (reported in the second row of Table \ref{tab:Table 3}). The correlation coefficients are 0.31 with a \textit{p}-value of 0.16 for the case of the molecular gas mass calculated within a 100~pc radius aperture, 0.29 with a \textit{p}-value of 0.21 for the 75~pc radius aperture, and 0.29 with a \textit{p}-value of 0.26 for the 50~pc radius aperture. 

\subsubsection{mm continuum luminosity-- molecular gas mass correlation}
\label{4.1.3}
\begin{figure*}
    \centering
    \includegraphics[width=18cm]{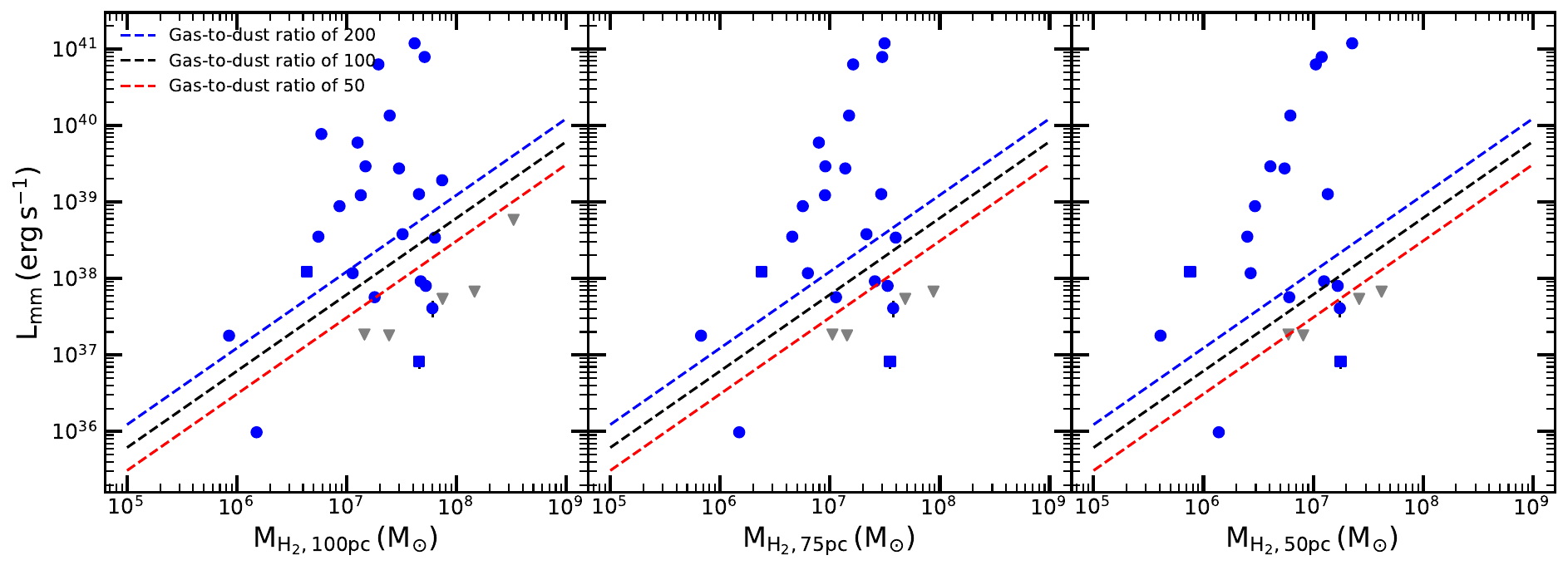}
    \caption{As Figure~\ref{fig:Figure 1}, but for the nuclear mm-continuum luminosity. Grey triangles are used for the upper limits (calculated as three times the rms noise level) of galaxies undetected in the mm continuum. 
    }
    \label{fig:Figure 3}
\end{figure*}

The nuclear millimetre continuum luminosity is another proxy of the nuclear activity. Indeed, excess mm luminosity has been observed in AGN hosting galaxies, with the excess being attributed to the AGN itself \citep[e.g.][]{2015MNRAS.451..517B,2018MNRAS.478..399B,2016PASJ...68...56D,2018ApJ...855...46W,2022arXiv220803880K}. We show in Figure~\ref{fig:Figure 3} the {total} nuclear millimetre luminosity (calculated on scales $\lesssim200$~pc) against the molecular gas mass on sub-kiloparsec scales, again finding no correlation between the two. 
The lack of correlation is supported by the Spearman rank analysis (reported in the third row of Table \ref{tab:Table 3}). The correlation coefficients are 0.002, 0.14 and 0.11 with \textit{p}-values 0.99, 0.55 and 0.66 for the 100~pc, 75~pc and 50~pc regions, respectively. 

\subsection{Accretion rate--mass correlation} 
\label{4.2}
\begin{figure*}
    \centering
    \includegraphics[width=18cm]{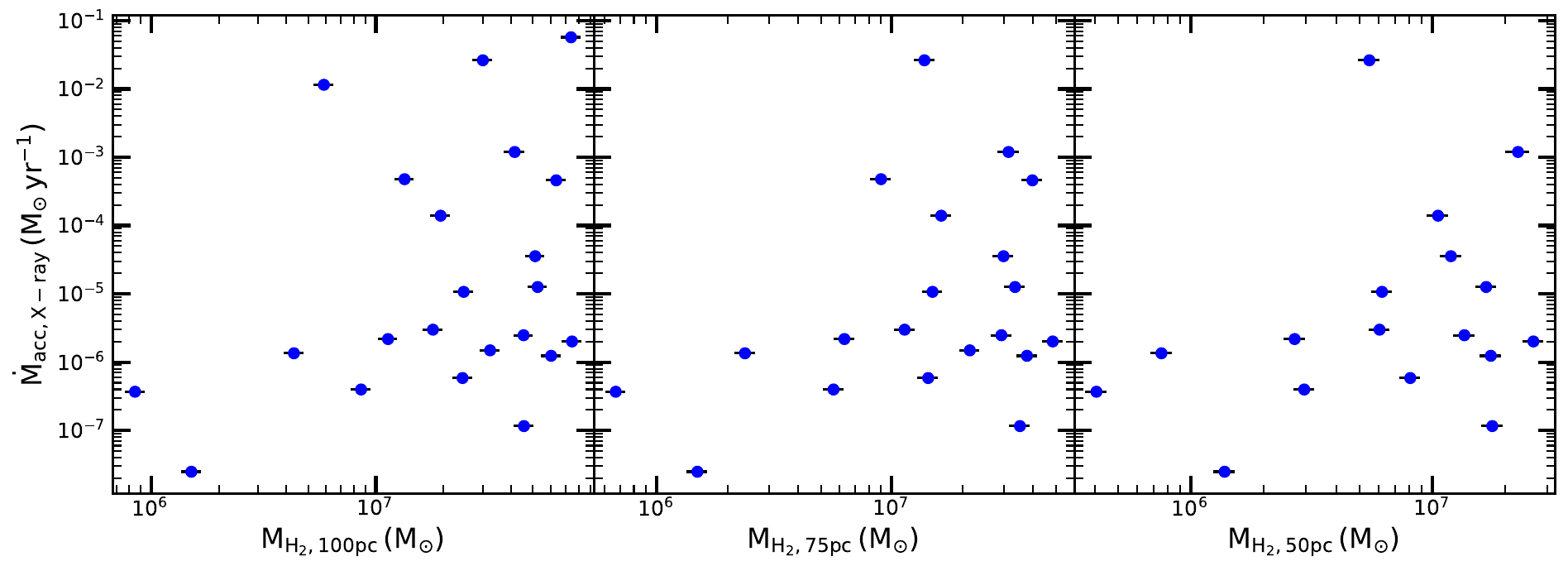}
    \caption{As Figure~\ref{fig:Figure 1}, but for the X-ray-traced accretion rate. 
    }
    \label{fig:Figure 4}
\end{figure*}

\begin{figure*}
    \centering
    \includegraphics[width=18cm]{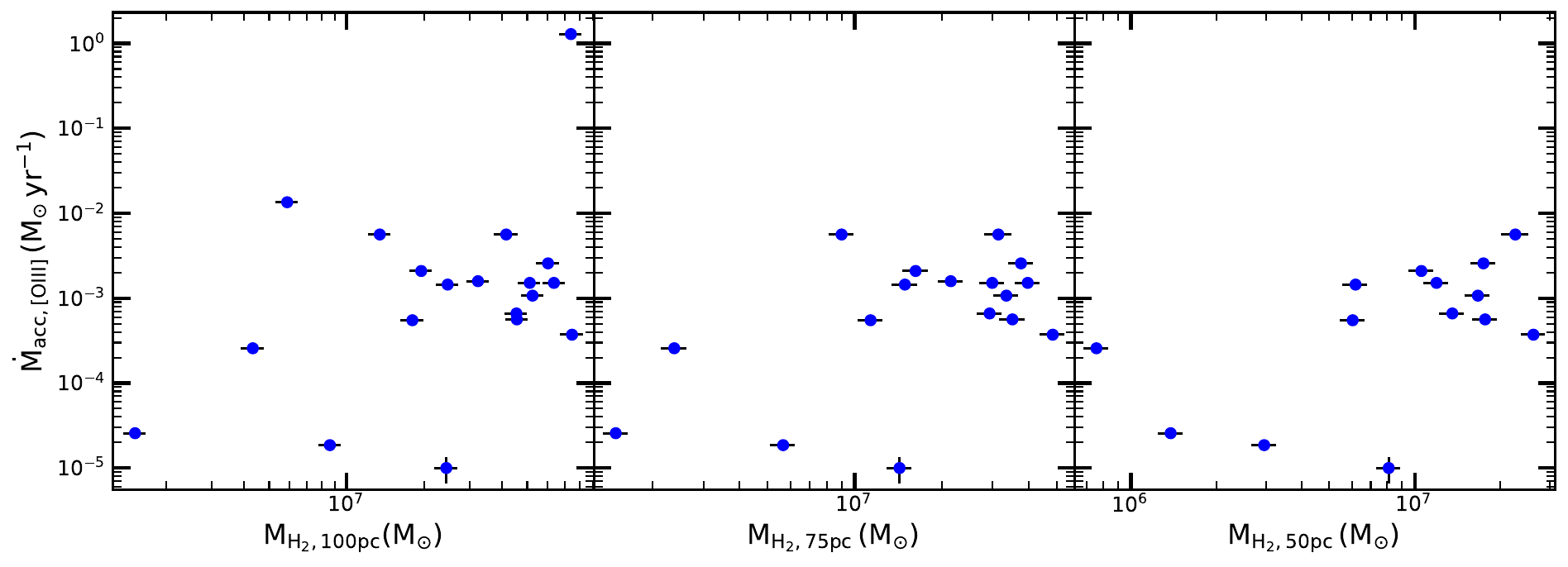}
    \caption{As Figure~\ref{fig:Figure 1}, but for the [OIII]-traced accretion rate. 
    }
    \label{fig:Figure 5}
\end{figure*}

In Figures~\ref{fig:Figure 4} and \ref{fig:Figure 5} we investigate relations between the molecular gas mass and AGN accretion rate, as calculated using the 2-10~keV and [OIII] line luminosity proxies. 
Also in this case, we do not find any clear correlation. The corresponding Spearman rank analysis (reported in the fourth and fifth rows of Table \ref{tab:Table 3}) mostly confirms this scenario.

For accretion rates calculated using the 2-10 keV luminosity as a proxy, a Spearman rank correlation coefficient of 0.31 with a \textit{p}-value of 0.16 is obtained for the 100~pc radius aperture. For the 75 and 50~pc regions, the Spearman rank coefficients are 0.29 and 0.29, respectively,  with associated \textit{p}-values of 0.21 and 0.26. 

For accretion rates calculated using the [OIII] line luminosity as a proxy, the Spearman rank coefficients are 0.27, 0.38 and 0.53 with p-values of 0.26, 0.13 and 0.05 for the 100~pc, 75~pc and 50~pc regions, respectively. These values imply that - as the aperture size decreases - a mildly significant correlation seems to be present. Whether this is real or coincidental due to the reduced number of data points should be investigated further.   

\section{Discussion} 
\label{Section 5}

\subsection{AGN activity and the circumnuclear molecular gas}
\label{5.1}

As described above, we do not find any strong correlation between the masses of molecular gas in the nuclear regions of our diverse sample galaxies and their AGN activities, as traced in a variety of ways. This is despite multiple authors reporting such correlations when studying specific AGN-selected samples of galaxies (\citealp[e.g][]{Izumi16,Babyk19,Koss22}). Our galaxies were not selected to be AGN, but do cover similar ranges of AGN properties (see Figure \ref{fig:Hists}).

Our results suggests that the level of nuclear activity in a given galaxy cannot purely depend on the amount of cold gas around its SMBH. This supports recent work by \protect\cite{2023arXiv230401017M}, where no correlation between the CO(2-1) and AGN luminosity has been reported. In this work, \protect \cite{2023arXiv230401017M} looked at the correlations between the cold molecular gas mass and AGN properties in a sample of radiative-mode AGN at z$\lesssim$0.5, using the 5100\AA\,AGN luminosity as analogy for the AGN bolometric luminosity. In this way, they initially find a weak correlation between CO(2-1) and AGN luminosity, but this then disappears when correcting for the cosmic evolution of the molecular gas content in galaxies. 
This also clearly highlights that the mechanism(s) driving gas from the wider galaxy scales to the nuclear regions are likely to be different in different types of AGN, and that timescale variations may be important. Below we discuss each of our tracers, and the conclusions that can be drawn from the differences between our results and others in literature.

\subsubsection{X-ray emission}

As described in Sections~\ref{4.1.2} and \ref{4.2} (and illustrated in Figures~\ref{fig:Figure 2} and \ref{fig:Figure 4}, respectively), there is no correlation between the cold molecular gas masses in the circumnuclear regions and the X-ray luminosities/X-ray derived accretion rates of our sample galaxies. {These results are consistent with those found in other recent works. \citet{2018MNRAS.473.5658R} looked at the relationship between the 2--10~keV X-ray luminosity and the \linebreak CO(2-1) brightness in a sample of Seyfert galaxies with X-ray luminosities that fall within the range of those studied in our work ($\approx 10^{41.5}$ to $\approx 10^{43.5} {\rm erg\,s^{-1}}$). The authors find no correlation between the two quantities, and ascribe that to the differences between the spatial scale probed by the single dish beam and and that of the SMBH accretion disc. \citet{2021A&A...652A..98G} also studied the correlation between the molecular gas mass on kpc scales (0.4-1.2\,kpc) and the X-ray luminosity in a sample of nearby Seyferts with similar $L_{X, 2-10})$ ($\approx 10^{39}$ to $\approx 10^{44} {\rm erg\,s^{-1}}$) and molecular gas masses ($\approx 10^{6.5}$ to $\approx 10^{9.5} {\rm M}_{\odot}$) to the objects studied in this work, finding again no correlation between the two quantities. In this case, the lack of correlation is explained by the different spatial scales and timescales involved with the last steps of the SMBH fuelling process and the kpc-scales molecular gas reservoirs.}\par
These results are in contrast to \cite{Izumi16}, who reported a positive correlation between the dense ($n_{\rm H_2} \gtrsim 10^{4-5} {\rm cm^{-3}}$) molecular gas mass at $\approx100$~pc scales and the X-ray traced accretion rates (calculated using the same methods adopted here) onto the SMBHs of a small sample of 10 nearby Seyfert galaxies. This was interpreted as supporting the role of CNDs in the AGN fuelling process.\par
Differences between their results and ours may arise for several reasons.
Firstly, as also discussed in \cite{2018MNRAS.473.5658R} and \cite{2021A&A...652A..98G} it is believed that in radiative-mode AGN X-ray emission traces recently accreted material, as in these cases X-rays are expected to be produced very close to the central SMBH \citep[e.g.][]{1979ApJ...229..318G}. In this scenario, X-ray emission can be highly time-variable. However, it would take dynamical timescales of several hundred thousand years for the gas at the scales we are investigating to fall onto the SMBHs. This difference in timescales and spatial scales could explain why we do not find correlation between the circumnuclear gas mass and the X-ray luminosity.\par
\citet{Izumi16} also investigated the correlation between dense molecular gas mass of the CND and accretion rate only in ten Seyfert galaxies, whereas we consider a $>3$ times larger sample of galaxies with a varied range of nuclear activities (see Figure \ref{fig:Hists}) and AGN types (see Table~\ref{tab:Table 1}). 
However, even when considering only the Seyferts in our sample, our results remain unchanged.\par
Furthermore, the correlation reported by \cite{Izumi16} involves the dense ($n_{\rm H_2} \gtrsim 10^{4-5} {\rm cm^{-3}}$) molecular gas mass of the CNDs, estimated using the HCN molecule as tracer. Here we instead use the {\it total} molecular gas mass on circumnuclear scales, which has been estimated via CO emission. However, we still cannot make the Seyfert galaxies in our sample follow the correlation of \cite{Izumi16} without requiring extremely low dense-gas fractions that also vary wildly between galaxies, a behaviour currently not observed in these kind of objects \citep[e.g.][]{2019ApJ...880..127J}.\par
We also note that the lack of correlation may be also ascribed to contamination from other (unresolved) sources of X-ray luminosity in the galaxies, such as stellar X-ray binaries (see Section~\ref{sec:2.3.1}). However, even when restricting our analysis only to sources observed at high spatial resolution with {\it Chandra} (i.e.\,where the nuclear emission from the AGN can be spatially isolated), we still do not observe any correlation. This suggests that contamination is not driving our results. 

Another more speculative possibility that could explain the lack of correlation is that different mechanisms usually give rise to the observed X-ray emission in different AGN types. In radiative-mode AGN such as Seyferts, X-rays are typically produced by inverse Compton up-scattering of photons from the accretion disc by the corona \citep{1997ApJ...487L.105C}. Whilst, in kinetic-mode AGN, classic accretion discs are either not present or truncated at inner radii (see Section~\ref{sec:Section1}), and X-ray emission likely arises from other processes, such as Compton up-scattering of non-thermal photons from the radio jets \citep[e.g.][]{2006ApJ...644L..13B}. These two emission processes may not correlate directly or may differ in how they correlate with the cold molecular gas mass of the circumnuclear regions.\par 
Finally another more speculative possibility is that the Seyfert galaxies observed by \cite{Izumi16} may have been caught in a special phase with bright HCN emission, possibly suggesting a bias in the sample selection. 
Such bright HCN emission may be more common in Seyferts with sizeable dense molecular gas reservoirs, leading also to higher accretion rates than most of those probed by our sample. Further investigation of diverse galaxy samples in central regions $<50$~pc in radius and using denser gas tracers will allow us to confirm/discard this hypothesis.



\subsubsection{Radio emission}



\cite{Babyk19} reported a correlation between the molecular gas mass up to kpc scales and jet power in a sample of nearby ETGs, most of which are LERGs. We wanted to expand this study probing the gas mass down to circumnuclear scales and in a more diverse sample of galaxies.
If the results reported by \cite{Babyk19} held at circumnuclear scales, we would have expected to observe at least some correlation between the circumnuclear gas masses of our sample galaxies and the excess 1.4~GHz radio luminosities (as $P_{\rm jet} \propto L_{1.4}$; e.g. \citealt{2010ApJ...720.1066C}). As discussed in Section \ref{4.1.1} and illustrated in Figure~\ref{fig:Figure 1}, we do not find any sign of such correlation in this work. This could be explained if the correlation reported by \cite{Babyk19} does not have anything to do with the SMBH fuelling, but instead arise because more massive galaxies tend to have more massive SMBHs, thus producing higher-power radio jets \citep{2006ApJ...637..669L}. {\cite{Babyk19} also report a link between the hot X-ray-emitting diffuse gas and the molecular gas content in their sample galaxies. This correlation is explained by the cooling of hot gas which is turned into molecular gas in the galaxy. This cooling of hot gas may then be connected to the radio power in these galaxies and would explain why we do not find a correlation at circumnuclear scales.}
We note that a result similar to that of \citet{Babyk19} has been recently reported by \cite{2023arXiv230316927F} and \cite{2023arXiv231003794F}, who studied the correlation between the jet power estimated from X-ray cavities ($P_{\rm cav}$) and the molecular gas mass within 500~pc in a sample of massive elliptical galaxies. This difference is results could be explained by the sample studied by these two works. In these works they study 9 and 13 objects respectively which is significantly smaller then the 35 objects used in our work. They also look for correlations exclusively in elliptical galaxies compared to diverse range of galaxies and activities types study in our work. 

Overall, the lack of correlation between the circumnuclear molecular gas mass and radio emission in our sample may suggest that jets are not directly powered by accretion from circumnuclear gas reservoirs, or that such correlation only exists over very long timescales. As discussed above, the dynamical times at the spatial scales probed here are still long compared to most AGN lifetime estimates. While radio jets can extend on large scales (and thus allow us to average AGN activity over longer timescales than direct tracers such as X-ray emission), this timescale mismatch may be too large to lead to any strong correlation. 


\subsubsection{Optical line emission}
As described in Section~\ref{sec:2.3.3}, we estimated the accretion rate in our sources using also the [OIII] line emission as a tracer, finding again no correlation with the molecular gas mass in the circumnuclear regions (see Section~\ref{4.2} and Figure~\ref{fig:Figure 5}). One could ascribe this lack of correlation to contamination from other sources of [OIII] emission on larger scales (e.g.\,[OIII] can also be a tracer of star-forming regions). This kind of contamination, however, has been found to be relevant only in higher redshift galaxies \citep[e.g.][]{2016MNRAS.462..181S}, and should therefore be minimal in nearby galaxies like our sample sources.

This result provides support to the hypotheses formulated above that either the level of nuclear activity in a given galaxy does not exclusively depend on the amount of cold gas around the central SMBH, or temporal variations in the accretion rate wash out any correlation. These results also support the idea that AGN fuelling mechanisms are not ubiquitous and different processes may be at play in different AGN types. 

   
\subsection{Nuclear activity and structure of the molecular gas reservoir}
\label{5.3}

For a sample of nearby Seyfert galaxies, \cite{2021A&A...652A..98G} reported that AGN luminosity (traced by 2-10~keV X-ray emission) correlates strongly with the structure of the molecular ISM (traced by CO) in the central 200\,pc. This could be due to AGN feedback impacting the cold molecular gas reservoirs at these scales, and driving the molecular gas away from (and/or heating/destroying it in) the centres of the galaxies.

\begin{figure}
    \centering
    \includegraphics[width=0.45\textwidth]{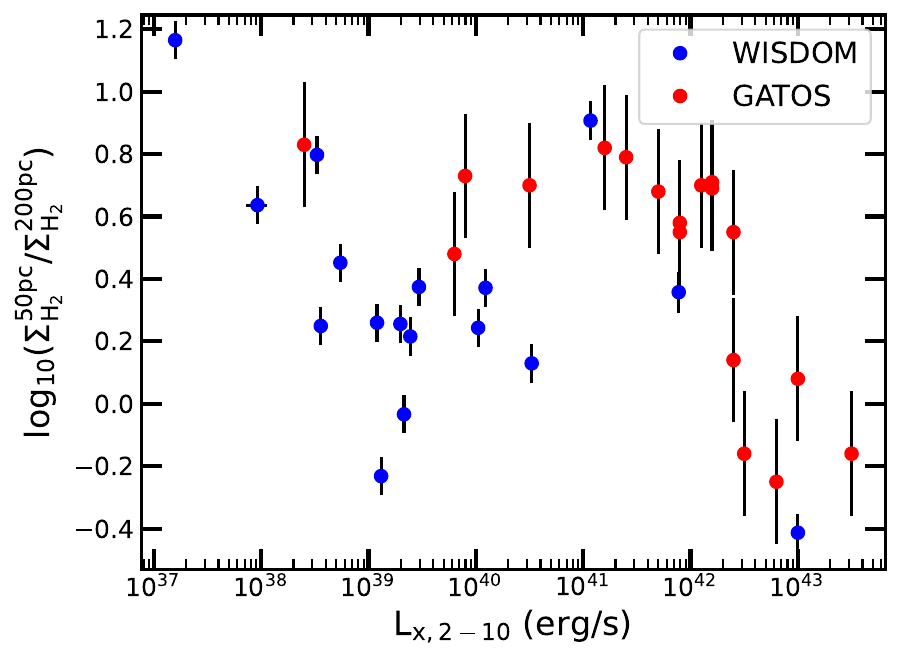}
    \caption{2-10~keV X-ray luminosity versus nuclear molecular gas concentration. Blue data points are for the sample analysed in this work, red data points are from \protect \cite{2021A&A...652A..98G}. The galaxies in our sample do not obey the trend reported in such previous work. This suggests that either the impact of AGN feedback is still detectable in a galaxy when it has gone into a lower activity phase, or the structure of the nuclear molecular gas is not determined by AGN processes.}
    \label{fig:Figure 10}
\end{figure}

As discussed in - e.g.\,- \cite{2018MNRAS.473.3818D}, some of the WISDOM sample galaxies have central molecular gas holes, and so it is possible the same mechanism is occurring here. We test this in Figure~\ref{fig:Figure 10}, where we plot the 2--10~keV X-ray luminosity against molecular gas concentration (as defined in Section \ref{Section 2}) for both our sample and that of \cite{2021A&A...652A..98G}. Our galaxies span a range of X-ray luminosities and molecular gas concentrations similar to those of \cite{2021A&A...652A..98G}, but do not seem to follow the same correlation.  

The lack of any correlation in our galaxy sample, which spans a wide range of AGN types and $L_{\rm bol}$ ($\rm 10^{41}\,-10^{46} \, erg \, s^{-1}$) and does include a significant number of Seyferts (albeit not selected to be especially active), suggests two possibilities:

\begin{enumerate}
\item The central structure of the molecular gas in galaxies is set by secular (non AGN-driven) process(es). The correlation of \cite{2021A&A...652A..98G} could then arise if these processes correlate with the SMBH mass (or another variable SMBH mass correlates with, such as spheroid mass/velocity dispersion), and thus the maximum AGN power possible. Some putative processes that could cause nuclear holes in the cold gas distributions (such as shear; see \citealt{2018MNRAS.473.3818D}) could naturally follow such a scaling. 
\item The central structure of the molecular gas in our galaxies has been impacted by AGN feedback, but the black hole is now in a phase of lower activity. If this was the case, a galaxy would be expected to obey the  \cite{2021A&A...652A..98G} correlation until its AGN episode dies off, then decrease in X-ray luminosity while presenting its feedback-affected molecular gas structure for some time, before further inflows reset the cycle. 
\end{enumerate}

{Comparing the WISDOM sample to the galaxies studied in \cite{2021A&A...652A..98G}, it seems that secular processes are the more likely scenario. Our sample contains more early-type hosts, but other galaxy and AGN properties are similar. It is unclear why the SMBH and its energy output (set on sub-parsec scales) would care about the large-scale galaxy properties, thus implying that the central molecular gas concentration is set by secular processes rather than by the nuclear activity. On the other hand, the parameter space explored in \cite{2021A&A...652A..98G} could represent the turnover point between secular processes and AGN feedback, with the turnover happening at X-ray luminosities around $10^{42}$ erg~s$^{-1}$. Our points would then fit this model with the sample AGN with X-ray luminosities lower than $10^{42}$ erg~s$^{-1}$ and a range of molecular gas concentrations set by secular processes. The few galaxies in our sample with X-ray luminosities higher than $10^{42}$ erg~s$^{-1}$ instead follow the relation found in \cite{2021A&A...652A..98G}.} Determining which of these possibilities is at work in our sample galaxies is interesting, but will require further observations and simulations of molecular gas at the centres of active galaxies of all luminosities. This will be explored further in future works.


\label{5.4}

\section{Conclusions} \label{Section 6}
We have searched for correlations between the cold molecular gas masses on the circumnuclear regions of a sample of 35 nearby galaxies and AGN activity tracers at radio, X-ray, optical and sub-mm wavelengths. We find that the molecular gas masses of our sample sources, measured within a range of elliptical apertures with radii from 50 to 100\,pc, do not correlate with any of the adopted tracers. 

The sample analysed in this study includes galaxies with a range of nuclear activities and global properties, and we are unable to reproduce any of the results found for other AGN-specific sub-samples. This suggests the level of nuclear activity in a given galaxy cannot purely be due the amount of cold gas fuel reservoir around the central SMBH. The fuelling mechanism of active galaxies is not ubiquitous and may vary between AGN types, and timescale variations are likely very important.

We also probed the molecular concentration of the circumnuclear gas discs in our sample galaxies to assess whether they had been impacted by AGN feedback. There is no evidence of a relation between structure on circumnuclear scales and current accretion rate, in contrast to results found for some nearby Seyfert galaxies selected to be in an active phase and despite our objects spanning the same range in circumnuclear properties.
This could indicate that these galaxies were previously in a more active phase that impacted the circumnuclear gas, or that these molecular concentrations arise naturally within circumnuclear gas discs and are not related to AGN processes. 
\par

Further observations and theoretical studies are clearly needed to make further progress to determine the link between circumnuclear gas reservoirs and nuclear activity. For instance, dense gas may be better linked to the direct reservoirs for accretion, and expanded sample sizes may help to overcome timescale issues. 

\section*{Acknowledgements}
This work is supported by the UKRI AIMLAC CDT, funded by grant EP/S023992/1. TAD and IR acknowledges support from STFC grant ST/S00033X/1. MB was supported by STFC consolidated grant "Astrophysics at Oxford" ST/H002456/1 and ST/K00106X/1. TGW acknowledges funding from the European Research Council (ERC) under the European Union’s Horizon 2020 research and innovation programme (grant agreement No. 694343). JG gratefully acknowledges financial support from the Swiss National Science Foundation (grant no CRSII5\_193826).
\par
This research made use of ASTROPY \footnote{\url{http://www.astropy.org/}},  a community-developed PYTHON package for Astronomy \citep{2013A&A...558A..33A,2018AJ....156..123A}, MATPLOTLIB \footnote{\url{https://matplotlib.org/}}, an open source visualisation package \citep{2007CSE.....9...90H}, NUMPY \footnote{\url{https://numpy.org/}}, an open source numerical computation library \citep{2020Natur.585..357H} and PANDAS\footnote{\url{https://pandas.pydata.org/}}, a data manipulation software library \citep{mckinney-proc-scipy-2010, the_pandas_development_team_2022_7093122}.
\par
This paper makes use of ALMA data. ALMA is a partnership of
the ESO (representing its member states), NSF (USA), and NINS
(Japan), together with the NRC (Canada), NSC, ASIAA (Taiwan), and KASI (Republic of Korea), in cooperation with the Republic
of Chile. The Joint ALMA Observatory is operated by the ESO,
AUI/NRAO, and NAOJ.
\par
This paper has also made use of the NASA/IPAC Extragalactic Database
(NED) which is operated by the Jet Propulsion Laboratory, California
Institute of Technology under contract with NASA.

\section*{Data Availability}
The data underlying this article are available in
the ALMA archive, at \url{http://almascience.eso.org/aq/}. Archival data are available from the NASA/IPAC Extragalactic Database (NED), at \url{https://ned.ipac.caltech.edu/} and Dustpedia, at \url{http://dustpedia.astro.noa.gr/}. The data used and the final plots will be shared upon a reasonable request to the first author.




\bibliographystyle{mnras}
\bibliography{example.bib}

\begin{thebibliography}{}
\makeatletter
\relax
\def\mn@urlcharsother{\let\do\@makeother \do\$\do\&\do\#\do\^\do\_\do\%\do\~}
\def\mn@doi{\begingroup\mn@urlcharsother \@ifnextchar [ {\mn@doi@}
  {\mn@doi@[]}}
\def\mn@doi@[#1]#2{\def\@tempa{#1}\ifx\@tempa\@empty \href
  {http://dx.doi.org/#2} {doi:#2}\else \href {http://dx.doi.org/#2} {#1}\fi
  \endgroup}
\def\mn@eprint#1#2{\mn@eprint@#1:#2::\@nil}
\def\mn@eprint@arXiv#1{\href {http://arxiv.org/abs/#1} {{\tt arXiv:#1}}}
\def\mn@eprint@dblp#1{\href {http://dblp.uni-trier.de/rec/bibtex/#1.xml}
  {dblp:#1}}
\def\mn@eprint@#1:#2:#3:#4\@nil{\def\@tempa {#1}\def\@tempb {#2}\def\@tempc
  {#3}\ifx \@tempc \@empty \let \@tempc \@tempb \let \@tempb \@tempa \fi \ifx
  \@tempb \@empty \def\@tempb {arXiv}\fi \@ifundefined
  {mn@eprint@\@tempb}{\@tempb:\@tempc}{\expandafter \expandafter \csname
  mn@eprint@\@tempb\endcsname \expandafter{\@tempc}}}

\bibitem[\protect\citeauthoryear{{Alexander} \& {Hickox}}{{Alexander} \&
  {Hickox}}{2012}]{2012NewAR..56...93A}
{Alexander} D.~M.,  {Hickox} R.~C.,  2012, \mn@doi [\nar]
  {10.1016/j.newar.2011.11.003}, \href
  {https://ui.adsabs.harvard.edu/abs/2012NewAR..56...93A} {56, 93}

\bibitem[\protect\citeauthoryear{{Allen}, {Dunn}, {Fabian}, {Taylor}  \&
  {Reynolds}}{{Allen} et~al.}{2006}]{2006MNRAS.372...21A}
{Allen} S.~W.,  {Dunn} R.~J.~H.,  {Fabian} A.~C.,  {Taylor} G.~B.,   {Reynolds}
  C.~S.,  2006, \mn@doi [\mnras] {10.1111/j.1365-2966.2006.10778.x}, \href
  {https://ui.adsabs.harvard.edu/abs/2006MNRAS.372...21A} {372, 21}

\bibitem[\protect\citeauthoryear{{Antonucci}}{{Antonucci}}{1993}]{Antonucci93}
{Antonucci} R.,  1993, \mn@doi [\araa] {10.1146/annurev.aa.31.090193.002353},
  \href {https://ui.adsabs.harvard.edu/abs/1993ARA&A..31..473A} {31, 473}

\bibitem[\protect\citeauthoryear{{Astropy Collaboration} et~al.,}{{Astropy
  Collaboration} et~al.}{2013}]{2013A&A...558A..33A}
{Astropy Collaboration} et~al., 2013, \mn@doi [\aap]
  {10.1051/0004-6361/201322068}, \href
  {https://ui.adsabs.harvard.edu/abs/2013A&A...558A..33A} {558, A33}

\bibitem[\protect\citeauthoryear{{Astropy Collaboration} et~al.,}{{Astropy
  Collaboration} et~al.}{2018}]{2018AJ....156..123A}
{Astropy Collaboration} et~al., 2018, \mn@doi [\aj] {10.3847/1538-3881/aabc4f},
  \href {https://ui.adsabs.harvard.edu/abs/2018AJ....156..123A} {156, 123}

\bibitem[\protect\citeauthoryear{{Babyk}, {McNamara}, {Tamhane}, {Nulsen},
  {Russell}  \& {Edge}}{{Babyk} et~al.}{2019}]{Babyk19}
{Babyk} I.~V.,  {McNamara} B.~R.,  {Tamhane} P.~D.,  {Nulsen} P.~E.~J.,
  {Russell} H.~R.,   {Edge} A.~C.,  2019, \mn@doi [\apj]
  {10.3847/1538-4357/ab54ce}, \href
  {https://ui.adsabs.harvard.edu/abs/2019ApJ...887..149B} {887, 149}

\bibitem[\protect\citeauthoryear{{Baldwin}, {Phillips}  \&
  {Terlevich}}{{Baldwin} et~al.}{1981}]{1981PASP...93....5B}
{Baldwin} J.~A.,  {Phillips} M.~M.,   {Terlevich} R.,  1981, \mn@doi [\pasp]
  {10.1086/130766}, \href
  {https://ui.adsabs.harvard.edu/abs/1981PASP...93....5B} {93, 5}

\bibitem[\protect\citeauthoryear{{Balmaverde}, {Baldi}  \&
  {Capetti}}{{Balmaverde} et~al.}{2008}]{2008A&A...486..119B}
{Balmaverde} B.,  {Baldi} R.~D.,   {Capetti} A.,  2008, \mn@doi [\aap]
  {10.1051/0004-6361:200809810}, \href
  {https://ui.adsabs.harvard.edu/abs/2008A&A...486..119B} {486, 119}

\bibitem[\protect\citeauthoryear{{Becker}, {White}  \& {Helfand}}{{Becker}
  et~al.}{1994}]{1994ASPC...61..165B}
{Becker} R.~H.,  {White} R.~L.,   {Helfand} D.~J.,  1994, in {Crabtree} D.~R.,
  {Hanisch} R.~J.,   {Barnes} J.,  eds,  Astronomical Society of the Pacific
  Conference Series Vol. 61, Astronomical Data Analysis Software and Systems
  III. p.~165

\bibitem[\protect\citeauthoryear{{Behar}, {Baldi}, {Laor}, {Horesh}, {Stevens}
  \& {Tzioumis}}{{Behar} et~al.}{2015}]{2015MNRAS.451..517B}
{Behar} E.,  {Baldi} R.~D.,  {Laor} A.,  {Horesh} A.,  {Stevens} J.,
  {Tzioumis} T.,  2015, \mn@doi [\mnras] {10.1093/mnras/stv988}, \href
  {https://ui.adsabs.harvard.edu/abs/2015MNRAS.451..517B} {451, 517}

\bibitem[\protect\citeauthoryear{{Behar}, {Vogel}, {Baldi}, {Smith}  \&
  {Mushotzky}}{{Behar} et~al.}{2018}]{2018MNRAS.478..399B}
{Behar} E.,  {Vogel} S.,  {Baldi} R.~D.,  {Smith} K.~L.,   {Mushotzky} R.~F.,
  2018, \mn@doi [\mnras] {10.1093/mnras/sty850}, \href
  {https://ui.adsabs.harvard.edu/abs/2018MNRAS.478..399B} {478, 399}

\bibitem[\protect\citeauthoryear{{Best} \& {Heckman}}{{Best} \&
  {Heckman}}{2012}]{2012MNRAS.421.1569B}
{Best} P.~N.,  {Heckman} T.~M.,  2012, \mn@doi [\mnras]
  {10.1111/j.1365-2966.2012.20414.x}, \href
  {https://ui.adsabs.harvard.edu/abs/2012MNRAS.421.1569B} {421, 1569}

\bibitem[\protect\citeauthoryear{{Bi}, {Feng}  \& {Ho}}{{Bi}
  et~al.}{2020}]{2020ApJ...900..124B}
{Bi} S.,  {Feng} H.,   {Ho} L.~C.,  2020, \mn@doi [\apj]
  {10.3847/1538-4357/aba761}, \href
  {https://ui.adsabs.harvard.edu/abs/2020ApJ...900..124B} {900, 124}

\bibitem[\protect\citeauthoryear{{Blundell}, {Fabian}, {Crawford}, {Erlund}  \&
  {Celotti}}{{Blundell} et~al.}{2006}]{2006ApJ...644L..13B}
{Blundell} K.~M.,  {Fabian} A.~C.,  {Crawford} C.~S.,  {Erlund} M.~C.,
  {Celotti} A.,  2006, \mn@doi [\apjl] {10.1086/504839}, \href
  {https://ui.adsabs.harvard.edu/abs/2006ApJ...644L..13B} {644, L13}

\bibitem[\protect\citeauthoryear{{Bolatto}, {Wolfire}  \& {Leroy}}{{Bolatto}
  et~al.}{2013}]{2013ARA&A..51..207B}
{Bolatto} A.~D.,  {Wolfire} M.,   {Leroy} A.~K.,  2013, \mn@doi [\araa]
  {10.1146/annurev-astro-082812-140944}, \href
  {https://ui.adsabs.harvard.edu/abs/2013ARA&A..51..207B} {51, 207}

\bibitem[\protect\citeauthoryear{{Bonatto} \& {Pastoriza}}{{Bonatto} \&
  {Pastoriza}}{1997}]{1997ApJ...486..132B}
{Bonatto} C.~J.,  {Pastoriza} M.~G.,  1997, \mn@doi [\apj] {10.1086/304511},
  \href {https://ui.adsabs.harvard.edu/abs/1997ApJ...486..132B} {486, 132}

\bibitem[\protect\citeauthoryear{{Bondi}}{{Bondi}}{1952}]{1952MNRAS.112..195B}
{Bondi} H.,  1952, \mn@doi [\mnras] {10.1093/mnras/112.2.195}, \href
  {https://ui.adsabs.harvard.edu/abs/1952MNRAS.112..195B} {112, 195}

\bibitem[\protect\citeauthoryear{{Boroson}, {Kim}  \& {Fabbiano}}{{Boroson}
  et~al.}{2011}]{2011ApJ...729...12B}
{Boroson} B.,  {Kim} D.-W.,   {Fabbiano} G.,  2011, \mn@doi [\apj]
  {10.1088/0004-637X/729/1/12}, \href
  {https://ui.adsabs.harvard.edu/abs/2011ApJ...729...12B} {729, 12}

\bibitem[\protect\citeauthoryear{{Bower}, {Benson}, {Malbon}, {Helly}, {Frenk},
  {Baugh}, {Cole}  \& {Lacey}}{{Bower} et~al.}{2006}]{2006MNRAS.370..645B}
{Bower} R.~G.,  {Benson} A.~J.,  {Malbon} R.,  {Helly} J.~C.,  {Frenk} C.~S.,
  {Baugh} C.~M.,  {Cole} S.,   {Lacey} C.~G.,  2006, \mn@doi [\mnras]
  {10.1111/j.1365-2966.2006.10519.x}, \href
  {https://ui.adsabs.harvard.edu/abs/2006MNRAS.370..645B} {370, 645}

\bibitem[\protect\citeauthoryear{{Buttiglione}, {Capetti}, {Celotti}, {Axon},
  {Chiaberge}, {Macchetto}  \& {Sparks}}{{Buttiglione}
  et~al.}{2009}]{2009A&A...495.1033B}
{Buttiglione} S.,  {Capetti} A.,  {Celotti} A.,  {Axon} D.~J.,  {Chiaberge} M.,
   {Macchetto} F.~D.,   {Sparks} W.~B.,  2009, \mn@doi [\aap]
  {10.1051/0004-6361:200811102}, \href
  {https://ui.adsabs.harvard.edu/abs/2009A&A...495.1033B} {495, 1033}

\bibitem[\protect\citeauthoryear{{Cappellari}}{{Cappellari}}{2013}]{2013ApJ...778L...2C}
{Cappellari} M.,  2013, \mn@doi [\apjl] {10.1088/2041-8205/778/1/L2}, \href
  {https://ui.adsabs.harvard.edu/abs/2013ApJ...778L...2C} {778, L2}

\bibitem[\protect\citeauthoryear{{Cappellari} et~al.,}{{Cappellari}
  et~al.}{2013}]{2013MNRAS.432.1862C}
{Cappellari} M.,  et~al., 2013, \mn@doi [\mnras] {10.1093/mnras/stt644}, \href
  {https://ui.adsabs.harvard.edu/abs/2013MNRAS.432.1862C} {432, 1862}

\bibitem[\protect\citeauthoryear{{Cavagnolo}, {McNamara}, {Nulsen}, {Carilli},
  {Jones}  \& {B{\^\i}rzan}}{{Cavagnolo} et~al.}{2010}]{2010ApJ...720.1066C}
{Cavagnolo} K.~W.,  {McNamara} B.~R.,  {Nulsen} P.~E.~J.,  {Carilli} C.~L.,
  {Jones} C.,   {B{\^\i}rzan} L.,  2010, \mn@doi [\apj]
  {10.1088/0004-637X/720/2/1066}, \href
  {https://ui.adsabs.harvard.edu/abs/2010ApJ...720.1066C} {720, 1066}

\bibitem[\protect\citeauthoryear{{Ciotti} \& {Ostriker}}{{Ciotti} \&
  {Ostriker}}{1997}]{1997ApJ...487L.105C}
{Ciotti} L.,  {Ostriker} J.~P.,  1997, \mn@doi [\apjl] {10.1086/310902}, \href
  {https://ui.adsabs.harvard.edu/abs/1997ApJ...487L.105C} {487, L105}

\bibitem[\protect\citeauthoryear{{Combes} et~al.,}{{Combes}
  et~al.}{2013}]{2013A&A...558A.124C}
{Combes} F.,  et~al., 2013, \mn@doi [\aap] {10.1051/0004-6361/201322288}, \href
  {https://ui.adsabs.harvard.edu/abs/2013A&A...558A.124C} {558, A124}

\bibitem[\protect\citeauthoryear{{Condon}, {Cotton}, {Greisen}, {Yin},
  {Perley}, {Taylor}  \& {Broderick}}{{Condon}
  et~al.}{1998}]{1998AJ....115.1693C}
{Condon} J.~J.,  {Cotton} W.~D.,  {Greisen} E.~W.,  {Yin} Q.~F.,  {Perley}
  R.~A.,  {Taylor} G.~B.,   {Broderick} J.~J.,  1998, \mn@doi [\aj]
  {10.1086/300337}, \href
  {https://ui.adsabs.harvard.edu/abs/1998AJ....115.1693C} {115, 1693}

\bibitem[\protect\citeauthoryear{{Cook}, {van Sistine}, {Singer}, {Kasliwal},
  {Kaplan}, {Iptf Collaboration}  \& {Growth Collaboration}}{{Cook}
  et~al.}{2017}]{2017GCN.21707....1C}
{Cook} D.~O.,  {van Sistine} A.,  {Singer} L.,  {Kasliwal} M.~M.,  {Kaplan} D.,
   {Iptf Collaboration}  {Growth Collaboration} 2017, GRB Coordinates Network,
  \href {https://ui.adsabs.harvard.edu/abs/2017GCN.21707....1C} {21707, 1}

\bibitem[\protect\citeauthoryear{{Crawford}, {Allen}, {Ebeling}, {Edge}  \&
  {Fabian}}{{Crawford} et~al.}{1999}]{1999MNRAS.306..857C}
{Crawford} C.~S.,  {Allen} S.~W.,  {Ebeling} H.,  {Edge} A.~C.,   {Fabian}
  A.~C.,  1999, \mn@doi [\mnras] {10.1046/j.1365-8711.1999.02583.x}, \href
  {https://ui.adsabs.harvard.edu/abs/1999MNRAS.306..857C} {306, 857}

\bibitem[\protect\citeauthoryear{{Croton} et~al.,}{{Croton}
  et~al.}{2006}]{2006MNRAS.365...11C}
{Croton} D.~J.,  et~al., 2006, \mn@doi [\mnras]
  {10.1111/j.1365-2966.2005.09675.x}, \href
  {https://ui.adsabs.harvard.edu/abs/2006MNRAS.365...11C} {365, 11}

\bibitem[\protect\citeauthoryear{{Davis}, {Bureau}, {Onishi}, {Cappellari},
  {Iguchi}  \& {Sarzi}}{{Davis} et~al.}{2017}]{2017MNRAS.468.4675D}
{Davis} T.~A.,  {Bureau} M.,  {Onishi} K.,  {Cappellari} M.,  {Iguchi} S.,
  {Sarzi} M.,  2017, \mn@doi [\mnras] {10.1093/mnras/stw3217}, \href
  {https://ui.adsabs.harvard.edu/abs/2017MNRAS.468.4675D} {468, 4675}

\bibitem[\protect\citeauthoryear{{Davis} et~al.,}{{Davis}
  et~al.}{2018}]{2018MNRAS.473.3818D}
{Davis} T.~A.,  et~al., 2018, \mn@doi [\mnras] {10.1093/mnras/stx2600}, \href
  {https://ui.adsabs.harvard.edu/abs/2018MNRAS.473.3818D} {473, 3818}

\bibitem[\protect\citeauthoryear{{Davis} et~al.,}{{Davis}
  et~al.}{2020}]{2020MNRAS.496.4061D}
{Davis} T.~A.,  et~al., 2020, \mn@doi [\mnras] {10.1093/mnras/staa1567}, \href
  {https://ui.adsabs.harvard.edu/abs/2020MNRAS.496.4061D} {496, 4061}

\bibitem[\protect\citeauthoryear{{Davis} et~al.,}{{Davis}
  et~al.}{2022}]{2022MNRAS.512.1522D}
{Davis} T.~A.,  et~al., 2022, \mn@doi [\mnras] {10.1093/mnras/stac600}, \href
  {https://ui.adsabs.harvard.edu/abs/2022MNRAS.512.1522D} {512, 1522}

\bibitem[\protect\citeauthoryear{{De Robertis} \& {Osterbrock}}{{De Robertis}
  \& {Osterbrock}}{1986}]{1986ApJ...301..727D}
{De Robertis} M.~M.,  {Osterbrock} D.~E.,  1986, \mn@doi [\apj]
  {10.1086/163939}, \href
  {https://ui.adsabs.harvard.edu/abs/1986ApJ...301..727D} {301, 727}

\bibitem[\protect\citeauthoryear{{Doi} \& {Inoue}}{{Doi} \&
  {Inoue}}{2016}]{2016PASJ...68...56D}
{Doi} A.,  {Inoue} Y.,  2016, \mn@doi [\pasj] {10.1093/pasj/psw052}, \href
  {https://ui.adsabs.harvard.edu/abs/2016PASJ...68...56D} {68, 56}

\bibitem[\protect\citeauthoryear{{Ferrarese} \& {Merritt}}{{Ferrarese} \&
  {Merritt}}{2000}]{2000ApJ...539L...9F}
{Ferrarese} L.,  {Merritt} D.,  2000, \mn@doi [\apjl] {10.1086/312838}, \href
  {https://ui.adsabs.harvard.edu/abs/2000ApJ...539L...9F} {539, L9}

\bibitem[\protect\citeauthoryear{{Fujita}, {Izumi}, {Kawakatu}, {Nagai},
  {Hirasawa}  \& {Ikeda}}{{Fujita} et~al.}{2023a}]{2023arXiv230316927F}
{Fujita} Y.,  {Izumi} T.,  {Kawakatu} N.,  {Nagai} H.,  {Hirasawa} R.,
  {Ikeda} Y.,  2023a, \mn@doi [arXiv e-prints] {10.48550/arXiv.2303.16927},
  \href {https://ui.adsabs.harvard.edu/abs/2023arXiv230316927F} {p.
  arXiv:2303.16927}

\bibitem[\protect\citeauthoryear{{Fujita}, {Izumi}, {Nagai}, {Kawakatu}  \&
  {Kawanaka}}{{Fujita} et~al.}{2023b}]{2023arXiv231003794F}
{Fujita} Y.,  {Izumi} T.,  {Nagai} H.,  {Kawakatu} N.,   {Kawanaka} N.,  2023b,
  \mn@doi [arXiv e-prints] {10.48550/arXiv.2310.03794}, \href
  {https://ui.adsabs.harvard.edu/abs/2023arXiv231003794F} {p. arXiv:2310.03794}

\bibitem[\protect\citeauthoryear{{Galeev}, {Rosner}  \& {Vaiana}}{{Galeev}
  et~al.}{1979}]{1979ApJ...229..318G}
{Galeev} A.~A.,  {Rosner} R.,   {Vaiana} G.~S.,  1979, \mn@doi [\apj]
  {10.1086/156957}, \href
  {https://ui.adsabs.harvard.edu/abs/1979ApJ...229..318G} {229, 318}

\bibitem[\protect\citeauthoryear{{Garc{\'\i}a-Burillo}
  et~al.,}{{Garc{\'\i}a-Burillo} et~al.}{2014}]{2014A&A...567A.125G}
{Garc{\'\i}a-Burillo} S.,  et~al., 2014, \mn@doi [\aap]
  {10.1051/0004-6361/201423843}, \href
  {https://ui.adsabs.harvard.edu/abs/2014A&A...567A.125G} {567, A125}

\bibitem[\protect\citeauthoryear{{Garc{\'\i}a-Burillo}
  et~al.,}{{Garc{\'\i}a-Burillo} et~al.}{2021}]{2021A&A...652A..98G}
{Garc{\'\i}a-Burillo} S.,  et~al., 2021, \mn@doi [\aap]
  {10.1051/0004-6361/202141075}, \href
  {https://ui.adsabs.harvard.edu/abs/2021A&A...652A..98G} {652, A98}

\bibitem[\protect\citeauthoryear{{Gaspari}, {Ruszkowski}  \& {Oh}}{{Gaspari}
  et~al.}{2013}]{2013MNRAS.432.3401G}
{Gaspari} M.,  {Ruszkowski} M.,   {Oh} S.~P.,  2013, \mn@doi [\mnras]
  {10.1093/mnras/stt692}, \href
  {https://ui.adsabs.harvard.edu/abs/2013MNRAS.432.3401G} {432, 3401}

\bibitem[\protect\citeauthoryear{{Gaspari}, {Brighenti}  \& {Temi}}{{Gaspari}
  et~al.}{2015}]{2015A&A...579A..62G}
{Gaspari} M.,  {Brighenti} F.,   {Temi} P.,  2015, \mn@doi [\aap]
  {10.1051/0004-6361/201526151}, \href
  {https://ui.adsabs.harvard.edu/abs/2015A&A...579A..62G} {579, A62}

\bibitem[\protect\citeauthoryear{{Gaspari}, {Temi}  \& {Brighenti}}{{Gaspari}
  et~al.}{2017}]{2017MNRAS.466..677G}
{Gaspari} M.,  {Temi} P.,   {Brighenti} F.,  2017, \mn@doi [\mnras]
  {10.1093/mnras/stw3108}, \href
  {https://ui.adsabs.harvard.edu/abs/2017MNRAS.466..677G} {466, 677}

\bibitem[\protect\citeauthoryear{{Gleisinger}, {O'Dea}, {Gallimore}, {Wykes}
  \& {Baum}}{{Gleisinger} et~al.}{2020}]{2020ApJ...905...42G}
{Gleisinger} R.~C.,  {O'Dea} C.~P.,  {Gallimore} J.~F.,  {Wykes} S.,   {Baum}
  S.~A.,  2020, \mn@doi [\apj] {10.3847/1538-4357/abc332}, \href
  {https://ui.adsabs.harvard.edu/abs/2020ApJ...905...42G} {905, 42}

\bibitem[\protect\citeauthoryear{{Grimm}, {Gilfanov}  \& {Sunyaev}}{{Grimm}
  et~al.}{2003}]{2003MNRAS.339..793G}
{Grimm} H.~J.,  {Gilfanov} M.,   {Sunyaev} R.,  2003, \mn@doi [\mnras]
  {10.1046/j.1365-8711.2003.06224.x}, \href
  {https://ui.adsabs.harvard.edu/abs/2003MNRAS.339..793G} {339, 793}

\bibitem[\protect\citeauthoryear{{G{\"u}ltekin} et~al.,}{{G{\"u}ltekin}
  et~al.}{2009}]{2009ApJ...698..198G}
{G{\"u}ltekin} K.,  et~al., 2009, \mn@doi [\apj] {10.1088/0004-637X/698/1/198},
  \href {https://ui.adsabs.harvard.edu/abs/2009ApJ...698..198G} {698, 198}

\bibitem[\protect\citeauthoryear{{Hardcastle}, {Evans}  \&
  {Croston}}{{Hardcastle} et~al.}{2007}]{2007MNRAS.376.1849H}
{Hardcastle} M.~J.,  {Evans} D.~A.,   {Croston} J.~H.,  2007, \mn@doi [\mnras]
  {10.1111/j.1365-2966.2007.11572.x}, \href
  {https://ui.adsabs.harvard.edu/abs/2007MNRAS.376.1849H} {376, 1849}

\bibitem[\protect\citeauthoryear{{Harris} et~al.,}{{Harris}
  et~al.}{2020}]{2020Natur.585..357H}
{Harris} C.~R.,  et~al., 2020, \mn@doi [\nat] {10.1038/s41586-020-2649-2},
  \href {https://ui.adsabs.harvard.edu/abs/2020Natur.585..357H} {585, 357}

\bibitem[\protect\citeauthoryear{{Harrison}}{{Harrison}}{2017}]{2017NatAs...1E.165H}
{Harrison} C.~M.,  2017, \mn@doi [Nature Astronomy] {10.1038/s41550-017-0165},
  \href {https://ui.adsabs.harvard.edu/abs/2017NatAs...1E.165H} {1, 0165}

\bibitem[\protect\citeauthoryear{{Heckman} \& {Best}}{{Heckman} \&
  {Best}}{2014}]{2014ARA&A..52..589H}
{Heckman} T.~M.,  {Best} P.~N.,  2014, \mn@doi [\araa]
  {10.1146/annurev-astro-081913-035722}, \href
  {https://ui.adsabs.harvard.edu/abs/2014ARA&A..52..589H} {52, 589}

\bibitem[\protect\citeauthoryear{{Heckman}, {Kauffmann}, {Brinchmann},
  {Charlot}, {Tremonti}  \& {White}}{{Heckman}
  et~al.}{2004}]{2004ApJ...613..109H}
{Heckman} T.~M.,  {Kauffmann} G.,  {Brinchmann} J.,  {Charlot} S.,  {Tremonti}
  C.,   {White} S. D.~M.,  2004, \mn@doi [\apj] {10.1086/422872}, \href
  {https://ui.adsabs.harvard.edu/abs/2004ApJ...613..109H} {613, 109}

\bibitem[\protect\citeauthoryear{{Ho}}{{Ho}}{2008}]{2008ARA&A..46..475H}
{Ho} L.~C.,  2008, \mn@doi [\araa] {10.1146/annurev.astro.45.051806.110546},
  \href {https://ui.adsabs.harvard.edu/abs/2008ARA&A..46..475H} {46, 475}

\bibitem[\protect\citeauthoryear{{Ho}, {Filippenko}  \& {Sargent}}{{Ho}
  et~al.}{1995}]{1995ApJS...98..477H}
{Ho} L.~C.,  {Filippenko} A.~V.,   {Sargent} W.~L.,  1995, \mn@doi [\apjs]
  {10.1086/192170}, \href
  {https://ui.adsabs.harvard.edu/abs/1995ApJS...98..477H} {98, 477}

\bibitem[\protect\citeauthoryear{{Hunter}}{{Hunter}}{2007}]{2007CSE.....9...90H}
{Hunter} J.~D.,  2007, \mn@doi [Computing in Science and Engineering]
  {10.1109/MCSE.2007.55}, \href
  {https://ui.adsabs.harvard.edu/abs/2007CSE.....9...90H} {9, 90}

\bibitem[\protect\citeauthoryear{{Izumi}, {Kawakatu}  \& {Kohno}}{{Izumi}
  et~al.}{2016}]{Izumi16}
{Izumi} T.,  {Kawakatu} N.,   {Kohno} K.,  2016, \mn@doi [\apj]
  {10.3847/0004-637X/827/1/81}, \href
  {https://ui.adsabs.harvard.edu/abs/2016ApJ...827...81I} {827, 81}

\bibitem[\protect\citeauthoryear{{Jarrett}, {Chester}, {Cutri}, {Schneider}  \&
  {Huchra}}{{Jarrett} et~al.}{2003}]{2003AJ....125..525J}
{Jarrett} T.~H.,  {Chester} T.,  {Cutri} R.,  {Schneider} S.~E.,   {Huchra}
  J.~P.,  2003, \mn@doi [\aj] {10.1086/345794}, \href
  {https://ui.adsabs.harvard.edu/abs/2003AJ....125..525J} {125, 525}

\bibitem[\protect\citeauthoryear{{Jim{\'e}nez-Donaire}
  et~al.,}{{Jim{\'e}nez-Donaire} et~al.}{2019}]{2019ApJ...880..127J}
{Jim{\'e}nez-Donaire} M.~J.,  et~al., 2019, \mn@doi [\apj]
  {10.3847/1538-4357/ab2b95}, \href
  {https://ui.adsabs.harvard.edu/abs/2019ApJ...880..127J} {880, 127}

\bibitem[\protect\citeauthoryear{{Kawamuro} et~al.,}{{Kawamuro}
  et~al.}{2022}]{2022arXiv220803880K}
{Kawamuro} T.,  et~al., 2022, arXiv e-prints, \href
  {https://ui.adsabs.harvard.edu/abs/2022arXiv220803880K} {p. arXiv:2208.03880}

\bibitem[\protect\citeauthoryear{{Kewley}, {Groves}, {Kauffmann}  \&
  {Heckman}}{{Kewley} et~al.}{2006}]{Kewley06}
{Kewley} L.~J.,  {Groves} B.,  {Kauffmann} G.,   {Heckman} T.,  2006, \mn@doi
  [\mnras] {10.1111/j.1365-2966.2006.10859.x}, \href
  {https://ui.adsabs.harvard.edu/abs/2006MNRAS.372..961K} {372, 961}

\bibitem[\protect\citeauthoryear{{Kim} \& {Fabbiano}}{{Kim} \&
  {Fabbiano}}{2004}]{2004ApJ...611..846K}
{Kim} D.-W.,  {Fabbiano} G.,  2004, \mn@doi [\apj] {10.1086/422210}, \href
  {https://ui.adsabs.harvard.edu/abs/2004ApJ...611..846K} {611, 846}

\bibitem[\protect\citeauthoryear{{King} \& {Nixon}}{{King} \&
  {Nixon}}{2015}]{2015MNRAS.453L..46K}
{King} A.,  {Nixon} C.,  2015, \mn@doi [\mnras] {10.1093/mnrasl/slv098}, \href
  {https://ui.adsabs.harvard.edu/abs/2015MNRAS.453L..46K} {453, L46}

\bibitem[\protect\citeauthoryear{{King} \& {Pounds}}{{King} \&
  {Pounds}}{2015}]{2015ARA&A..53..115K}
{King} A.,  {Pounds} K.,  2015, \mn@doi [\araa]
  {10.1146/annurev-astro-082214-122316}, \href
  {https://ui.adsabs.harvard.edu/abs/2015ARA&A..53..115K} {53, 115}

\bibitem[\protect\citeauthoryear{{King} \& {Pringle}}{{King} \&
  {Pringle}}{2007}]{2007MNRAS.377L..25K}
{King} A.~R.,  {Pringle} J.~E.,  2007, \mn@doi [\mnras]
  {10.1111/j.1745-3933.2007.00296.x}, \href
  {https://ui.adsabs.harvard.edu/abs/2007MNRAS.377L..25K} {377, L25}

\bibitem[\protect\citeauthoryear{{Komissarov} \& {Gubanov}}{{Komissarov} \&
  {Gubanov}}{1994}]{1994A&A...285...27K}
{Komissarov} S.~S.,  {Gubanov} A.~G.,  1994, \aap, \href
  {https://ui.adsabs.harvard.edu/abs/1994A&A...285...27K} {285, 27}

\bibitem[\protect\citeauthoryear{{Kormendy} \& {Ho}}{{Kormendy} \&
  {Ho}}{2013}]{2013ARA&A..51..511K}
{Kormendy} J.,  {Ho} L.~C.,  2013, \mn@doi [\araa]
  {10.1146/annurev-astro-082708-101811}, \href
  {https://ui.adsabs.harvard.edu/abs/2013ARA&A..51..511K} {51, 511}

\bibitem[\protect\citeauthoryear{{Koss} et~al.,}{{Koss} et~al.}{2021}]{Koss22}
{Koss} M.~J.,  et~al., 2021, \mn@doi [\apjs] {10.3847/1538-4365/abcbfe}, \href
  {https://ui.adsabs.harvard.edu/abs/2021ApJS..252...29K} {252, 29}

\bibitem[\protect\citeauthoryear{{Laing} \& {Bridle}}{{Laing} \&
  {Bridle}}{2013}]{2013MNRAS.432.1114L}
{Laing} R.~A.,  {Bridle} A.~H.,  2013, \mn@doi [\mnras] {10.1093/mnras/stt531},
  \href {https://ui.adsabs.harvard.edu/abs/2013MNRAS.432.1114L} {432, 1114}

\bibitem[\protect\citeauthoryear{{Lelli}, {Davis}, {Bureau}, {Cappellari},
  {Liu}, {Ruffa}, {Smith}  \& {Williams}}{{Lelli}
  et~al.}{2022}]{2022MNRAS.516.4066L}
{Lelli} F.,  {Davis} T.~A.,  {Bureau} M.,  {Cappellari} M.,  {Liu} L.,  {Ruffa}
  I.,  {Smith} M.~D.,   {Williams} T.~G.,  2022, \mn@doi [\mnras]
  {10.1093/mnras/stac2493}, \href
  {https://ui.adsabs.harvard.edu/abs/2022MNRAS.516.4066L} {516, 4066}

\bibitem[\protect\citeauthoryear{{Leroy}, {Walter}, {Brinks}, {Bigiel}, {de
  Blok}, {Madore}  \& {Thornley}}{{Leroy} et~al.}{2008}]{2008AJ....136.2782L}
{Leroy} A.~K.,  {Walter} F.,  {Brinks} E.,  {Bigiel} F.,  {de Blok} W.~J.~G.,
  {Madore} B.,   {Thornley} M.~D.,  2008, \mn@doi [\aj]
  {10.1088/0004-6256/136/6/2782}, \href
  {https://ui.adsabs.harvard.edu/abs/2008AJ....136.2782L} {136, 2782}

\bibitem[\protect\citeauthoryear{{Leroy} et~al.,}{{Leroy}
  et~al.}{2019}]{2019ApJS..244...24L}
{Leroy} A.~K.,  et~al., 2019, \mn@doi [\apjs] {10.3847/1538-4365/ab3925}, \href
  {https://ui.adsabs.harvard.edu/abs/2019ApJS..244...24L} {244, 24}

\bibitem[\protect\citeauthoryear{{Leroy} et~al.,}{{Leroy}
  et~al.}{2022}]{2022ApJ...927..149L}
{Leroy} A.~K.,  et~al., 2022, \mn@doi [\apj] {10.3847/1538-4357/ac3490}, \href
  {https://ui.adsabs.harvard.edu/abs/2022ApJ...927..149L} {927, 149}

\bibitem[\protect\citeauthoryear{{Liu}, {Jiang}  \& {Gu}}{{Liu}
  et~al.}{2006}]{2006ApJ...637..669L}
{Liu} Y.,  {Jiang} D.~R.,   {Gu} M.~F.,  2006, \mn@doi [\apj] {10.1086/498639},
  \href {https://ui.adsabs.harvard.edu/abs/2006ApJ...637..669L} {637, 669}

\bibitem[\protect\citeauthoryear{{Maccagni} et~al.,}{{Maccagni}
  et~al.}{2023}]{Maccagni23}
{Maccagni} F.~M.,  et~al., 2023, \mn@doi [\aap] {10.1051/0004-6361/202346521},
  \href {https://ui.adsabs.harvard.edu/abs/2023A&A...675A..59M} {675, A59}

\bibitem[\protect\citeauthoryear{{Magorrian} et~al.,}{{Magorrian}
  et~al.}{1998}]{1998AJ....115.2285M}
{Magorrian} J.,  et~al., 1998, \mn@doi [\aj] {10.1086/300353}, \href
  {https://ui.adsabs.harvard.edu/abs/1998AJ....115.2285M} {115, 2285}

\bibitem[\protect\citeauthoryear{{Marconi} \& {Hunt}}{{Marconi} \&
  {Hunt}}{2003}]{2003ApJ...589L..21M}
{Marconi} A.,  {Hunt} L.~K.,  2003, \mn@doi [\apjl] {10.1086/375804}, \href
  {https://ui.adsabs.harvard.edu/abs/2003ApJ...589L..21M} {589, L21}

\bibitem[\protect\citeauthoryear{{Marconi}, {Risaliti}, {Gilli}, {Hunt},
  {Maiolino}  \& {Salvati}}{{Marconi} et~al.}{2004}]{2004MNRAS.351..169M}
{Marconi} A.,  {Risaliti} G.,  {Gilli} R.,  {Hunt} L.~K.,  {Maiolino} R.,
  {Salvati} M.,  2004, \mn@doi [\mnras] {10.1111/j.1365-2966.2004.07765.x},
  \href {https://ui.adsabs.harvard.edu/abs/2004MNRAS.351..169M} {351, 169}

\bibitem[\protect\citeauthoryear{{Martini}, {Regan}, {Mulchaey}  \&
  {Pogge}}{{Martini} et~al.}{2003}]{Martini03}
{Martini} P.,  {Regan} M.~W.,  {Mulchaey} J.~S.,   {Pogge} R.~W.,  2003,
  \mn@doi [\apj] {10.1086/374685}, \href
  {https://ui.adsabs.harvard.edu/abs/2003ApJ...589..774M} {589, 774}

\bibitem[\protect\citeauthoryear{{McMullin}, {Waters}, {Schiebel}, {Young}  \&
  {Golap}}{{McMullin} et~al.}{2007}]{2007ASPC..376..127M}
{McMullin} J.~P.,  {Waters} B.,  {Schiebel} D.,  {Young} W.,   {Golap} K.,
  2007, in {Shaw} R.~A.,  {Hill} F.,   {Bell} D.~J.,  eds,  Astronomical
  Society of the Pacific Conference Series Vol. 376, Astronomical Data Analysis
  Software and Systems XVI. p.~127

\bibitem[\protect\citeauthoryear{{McNamara}, {Russell}, {Nulsen}, {Hogan},
  {Fabian}, {Pulido}  \& {Edge}}{{McNamara} et~al.}{2016}]{2016ApJ...830...79M}
{McNamara} B.~R.,  {Russell} H.~R.,  {Nulsen} P.~E.~J.,  {Hogan} M.~T.,
  {Fabian} A.~C.,  {Pulido} F.,   {Edge} A.~C.,  2016, \mn@doi [\apj]
  {10.3847/0004-637X/830/2/79}, \href
  {https://ui.adsabs.harvard.edu/abs/2016ApJ...830...79M} {830, 79}

\bibitem[\protect\citeauthoryear{{Molina}, {Shangguan}, {Wang}, {Ho}, {Bauer}
  \& {Treister}}{{Molina} et~al.}{2023}]{2023arXiv230401017M}
{Molina} J.,  {Shangguan} J.,  {Wang} R.,  {Ho} L.~C.,  {Bauer} F.~E.,
  {Treister} E.,  2023, \mn@doi [arXiv e-prints] {10.48550/arXiv.2304.01017},
  \href {https://ui.adsabs.harvard.edu/abs/2023arXiv230401017M} {p.
  arXiv:2304.01017}

\bibitem[\protect\citeauthoryear{{Morganti}}{{Morganti}}{2017}]{2017FrASS...4...42M}
{Morganti} R.,  2017, \mn@doi [Frontiers in Astronomy and Space Sciences]
  {10.3389/fspas.2017.00042}, \href
  {https://ui.adsabs.harvard.edu/abs/2017FrASS...4...42M} {4, 42}

\bibitem[\protect\citeauthoryear{{Moustakas}, {Kennicutt}, {Tremonti}, {Dale},
  {Smith}  \& {Calzetti}}{{Moustakas} et~al.}{2010}]{2010ApJS..190..233M}
{Moustakas} J.,  {Kennicutt} Robert~C. J.,  {Tremonti} C.~A.,  {Dale} D.~A.,
  {Smith} J.-D.~T.,   {Calzetti} D.,  2010, \mn@doi [\apjs]
  {10.1088/0067-0049/190/2/233}, \href
  {https://ui.adsabs.harvard.edu/abs/2010ApJS..190..233M} {190, 233}

\bibitem[\protect\citeauthoryear{{Murphy} et~al.,}{{Murphy}
  et~al.}{2011}]{2011ApJ...737...67M}
{Murphy} E.~J.,  et~al., 2011, \mn@doi [\apj] {10.1088/0004-637X/737/2/67},
  \href {https://ui.adsabs.harvard.edu/abs/2011ApJ...737...67M} {737, 67}

\bibitem[\protect\citeauthoryear{{Narayan} \& {Yi}}{{Narayan} \&
  {Yi}}{1995}]{1995ApJ...452..710N}
{Narayan} R.,  {Yi} I.,  1995, \mn@doi [\apj] {10.1086/176343}, \href
  {https://ui.adsabs.harvard.edu/abs/1995ApJ...452..710N} {452, 710}

\bibitem[\protect\citeauthoryear{{Nayakshin}, {Power}  \& {King}}{{Nayakshin}
  et~al.}{2012}]{2012ApJ...753...15N}
{Nayakshin} S.,  {Power} C.,   {King} A.~R.,  2012, \mn@doi [\apj]
  {10.1088/0004-637X/753/1/15}, \href
  {https://ui.adsabs.harvard.edu/abs/2012ApJ...753...15N} {753, 15}

\bibitem[\protect\citeauthoryear{{Negri}, {Posacki}, {Pellegrini}  \&
  {Ciotti}}{{Negri} et~al.}{2014}]{Negri14}
{Negri} A.,  {Posacki} S.,  {Pellegrini} S.,   {Ciotti} L.,  2014, \mn@doi
  [\mnras] {10.1093/mnras/stu1834}, \href
  {https://ui.adsabs.harvard.edu/abs/2014MNRAS.445.1351N} {445, 1351}

\bibitem[\protect\citeauthoryear{{North} et~al.,}{{North}
  et~al.}{2019}]{North19}
{North} E.~V.,  et~al., 2019, \mn@doi [\mnras] {10.1093/mnras/stz2598}, \href
  {https://ui.adsabs.harvard.edu/abs/2019MNRAS.490..319N} {490, 319}

\bibitem[\protect\citeauthoryear{{North} et~al.,}{{North}
  et~al.}{2021}]{2021MNRAS.503.5179N}
{North} E.~V.,  et~al., 2021, \mn@doi [\mnras] {10.1093/mnras/stab793}, \href
  {https://ui.adsabs.harvard.edu/abs/2021MNRAS.503.5179N} {503, 5179}

\bibitem[\protect\citeauthoryear{{Oca{\~n}a Flaquer}, {Leon}, {Combes}  \&
  {Lim}}{{Oca{\~n}a Flaquer} et~al.}{2010}]{2010A&A...518A...9O}
{Oca{\~n}a Flaquer} B.,  {Leon} S.,  {Combes} F.,   {Lim} J.,  2010, \mn@doi
  [\aap] {10.1051/0004-6361/200913392}, \href
  {https://ui.adsabs.harvard.edu/abs/2010A&A...518A...9O} {518, A9}

\bibitem[\protect\citeauthoryear{{Onishi}, {Iguchi}, {Davis}, {Bureau},
  {Cappellari}, {Sarzi}  \& {Blitz}}{{Onishi}
  et~al.}{2017}]{2017MNRAS.468.4663O}
{Onishi} K.,  {Iguchi} S.,  {Davis} T.~A.,  {Bureau} M.,  {Cappellari} M.,
  {Sarzi} M.,   {Blitz} L.,  2017, \mn@doi [\mnras] {10.1093/mnras/stx631},
  \href {https://ui.adsabs.harvard.edu/abs/2017MNRAS.468.4663O} {468, 4663}

\bibitem[\protect\citeauthoryear{{Pizzolato} \& {Soker}}{{Pizzolato} \&
  {Soker}}{2005}]{2005ApJ...632..821P}
{Pizzolato} F.,  {Soker} N.,  2005, \mn@doi [\apj] {10.1086/444344}, \href
  {https://ui.adsabs.harvard.edu/abs/2005ApJ...632..821P} {632, 821}

\bibitem[\protect\citeauthoryear{{Pizzolato} \& {Soker}}{{Pizzolato} \&
  {Soker}}{2010}]{2010MNRAS.408..961P}
{Pizzolato} F.,  {Soker} N.,  2010, \mn@doi [\mnras]
  {10.1111/j.1365-2966.2010.17156.x}, \href
  {https://ui.adsabs.harvard.edu/abs/2010MNRAS.408..961P} {408, 961}

\bibitem[\protect\citeauthoryear{{Prandoni}, {Laing}, {de Ruiter}  \&
  {Parma}}{{Prandoni} et~al.}{2010}]{2010A&A...523A..38P}
{Prandoni} I.,  {Laing} R.~A.,  {de Ruiter} H.~R.,   {Parma} P.,  2010, \mn@doi
  [\aap] {10.1051/0004-6361/201015456}, \href
  {https://ui.adsabs.harvard.edu/abs/2010A&A...523A..38P} {523, A38}

\bibitem[\protect\citeauthoryear{{Reeves} \& {Turner}}{{Reeves} \&
  {Turner}}{2000}]{2000MNRAS.316..234R}
{Reeves} J.~N.,  {Turner} M.~J.~L.,  2000, \mn@doi [\mnras]
  {10.1046/j.1365-8711.2000.03510.x}, \href
  {https://ui.adsabs.harvard.edu/abs/2000MNRAS.316..234R} {316, 234}

\bibitem[\protect\citeauthoryear{{Rodr{\'\i}guez-Ardila}, {Pastoriza}  \&
  {Donzelli}}{{Rodr{\'\i}guez-Ardila} et~al.}{2000}]{2000ApJS..126...63R}
{Rodr{\'\i}guez-Ardila} A.,  {Pastoriza} M.~G.,   {Donzelli} C.~J.,  2000,
  \mn@doi [\apjs] {10.1086/313293}, \href
  {https://ui.adsabs.harvard.edu/abs/2000ApJS..126...63R} {126, 63}

\bibitem[\protect\citeauthoryear{{Rosario} et~al.,}{{Rosario}
  et~al.}{2018}]{2018MNRAS.473.5658R}
{Rosario} D.~J.,  et~al., 2018, \mn@doi [\mnras] {10.1093/mnras/stx2670}, \href
  {https://ui.adsabs.harvard.edu/abs/2018MNRAS.473.5658R} {473, 5658}

\bibitem[\protect\citeauthoryear{{Ruffa} et~al.,}{{Ruffa}
  et~al.}{2019a}]{2019MNRAS.484.4239R}
{Ruffa} I.,  et~al., 2019a, \mn@doi [\mnras] {10.1093/mnras/stz255}, \href
  {https://ui.adsabs.harvard.edu/abs/2019MNRAS.484.4239R} {484, 4239}

\bibitem[\protect\citeauthoryear{{Ruffa} et~al.,}{{Ruffa}
  et~al.}{2019b}]{2019MNRAS.489.3739R}
{Ruffa} I.,  et~al., 2019b, \mn@doi [\mnras] {10.1093/mnras/stz2368}, \href
  {https://ui.adsabs.harvard.edu/abs/2019MNRAS.489.3739R} {489, 3739}

\bibitem[\protect\citeauthoryear{{Ruffa}, {Laing}, {Prandoni}, {Paladino},
  {Parma}, {Davis}  \& {Bureau}}{{Ruffa} et~al.}{2020}]{Ruffa20}
{Ruffa} I.,  {Laing} R.~A.,  {Prandoni} I.,  {Paladino} R.,  {Parma} P.,
  {Davis} T.~A.,   {Bureau} M.,  2020, \mn@doi [\mnras]
  {10.1093/mnras/staa3166}, \href
  {https://ui.adsabs.harvard.edu/abs/2020MNRAS.499.5719R} {499, 5719}

\bibitem[\protect\citeauthoryear{{Ruffa}, {Prandoni}, {Davis}, {Laing},
  {Paladino}, {Casasola}, {Parma}  \& {Bureau}}{{Ruffa}
  et~al.}{2022}]{2022MNRAS.510.4485R}
{Ruffa} I.,  {Prandoni} I.,  {Davis} T.~A.,  {Laing} R.~A.,  {Paladino} R.,
  {Casasola} V.,  {Parma} P.,   {Bureau} M.,  2022, \mn@doi [\mnras]
  {10.1093/mnras/stab3541}, \href
  {https://ui.adsabs.harvard.edu/abs/2022MNRAS.510.4485R} {510, 4485}

\bibitem[\protect\citeauthoryear{{Ruffa} et~al.,}{{Ruffa}
  et~al.}{2023a}]{Ruffa23b}
{Ruffa} I.,  et~al., 2023a, \mn@doi [arXiv e-prints]
  {10.48550/arXiv.2307.13872}, \href
  {https://ui.adsabs.harvard.edu/abs/2023arXiv230713872R} {p. arXiv:2307.13872}

\bibitem[\protect\citeauthoryear{{Ruffa} et~al.,}{{Ruffa}
  et~al.}{2023b}]{2023MNRAS.522.6170R}
{Ruffa} I.,  et~al., 2023b, \mn@doi [\mnras] {10.1093/mnras/stad1119}, \href
  {https://ui.adsabs.harvard.edu/abs/2023MNRAS.522.6170R} {522, 6170}

\bibitem[\protect\citeauthoryear{{Sandstrom} et~al.,}{{Sandstrom}
  et~al.}{2013}]{2013ApJ...777....5S}
{Sandstrom} K.~M.,  et~al., 2013, \mn@doi [\apj] {10.1088/0004-637X/777/1/5},
  \href {https://ui.adsabs.harvard.edu/abs/2013ApJ...777....5S} {777, 5}

\bibitem[\protect\citeauthoryear{{Seth} et~al.,}{{Seth}
  et~al.}{2010}]{2010ApJ...714..713S}
{Seth} A.~C.,  et~al., 2010, \mn@doi [\apj] {10.1088/0004-637X/714/1/713},
  \href {https://ui.adsabs.harvard.edu/abs/2010ApJ...714..713S} {714, 713}

\bibitem[\protect\citeauthoryear{{Shakura} \& {Sunyaev}}{{Shakura} \&
  {Sunyaev}}{1973}]{1973A&A....24..337S}
{Shakura} N.~I.,  {Sunyaev} R.~A.,  1973, \aap, \href
  {https://ui.adsabs.harvard.edu/abs/1973A&A....24..337S} {24, 337}

\bibitem[\protect\citeauthoryear{{Shlosman}, {Frank}  \& {Begelman}}{{Shlosman}
  et~al.}{1989}]{1989Natur.338...45S}
{Shlosman} I.,  {Frank} J.,   {Begelman} M.~C.,  1989, \mn@doi [\nat]
  {10.1038/338045a0}, \href
  {https://ui.adsabs.harvard.edu/abs/1989Natur.338...45S} {338, 45}

\bibitem[\protect\citeauthoryear{{Smith} et~al.,}{{Smith}
  et~al.}{2019}]{2019MNRAS.485.4359S}
{Smith} M.~D.,  et~al., 2019, \mn@doi [\mnras] {10.1093/mnras/stz625}, \href
  {https://ui.adsabs.harvard.edu/abs/2019MNRAS.485.4359S} {485, 4359}

\bibitem[\protect\citeauthoryear{{Smith} et~al.,}{{Smith}
  et~al.}{2021a}]{2021MNRAS.500.1933S}
{Smith} M.~D.,  et~al., 2021a, \mn@doi [\mnras] {10.1093/mnras/staa3274}, \href
  {https://ui.adsabs.harvard.edu/abs/2021MNRAS.500.1933S} {500, 1933}

\bibitem[\protect\citeauthoryear{{Smith} et~al.,}{{Smith}
  et~al.}{2021b}]{2021MNRAS.503.5984S}
{Smith} M.~D.,  et~al., 2021b, \mn@doi [\mnras] {10.1093/mnras/stab791}, \href
  {https://ui.adsabs.harvard.edu/abs/2021MNRAS.503.5984S} {503, 5984}

\bibitem[\protect\citeauthoryear{{Strong} et~al.,}{{Strong}
  et~al.}{1988}]{1988A&A...207....1S}
{Strong} A.~W.,  et~al., 1988, \aap, \href
  {https://ui.adsabs.harvard.edu/abs/1988A&A...207....1S} {207, 1}

\bibitem[\protect\citeauthoryear{{Suzuki} et~al.,}{{Suzuki}
  et~al.}{2016}]{2016MNRAS.462..181S}
{Suzuki} T.~L.,  et~al., 2016, \mn@doi [\mnras] {10.1093/mnras/stw1655}, \href
  {https://ui.adsabs.harvard.edu/abs/2016MNRAS.462..181S} {462, 181}

\bibitem[\protect\citeauthoryear{{Tadhunter}, {Morganti}, {di Serego
  Alighieri}, {Fosbury}  \& {Danziger}}{{Tadhunter}
  et~al.}{1993}]{1993MNRAS.263..999T}
{Tadhunter} C.~N.,  {Morganti} R.,  {di Serego Alighieri} S.,  {Fosbury}
  R.~A.~E.,   {Danziger} I.~J.,  1993, \mn@doi [\mnras]
  {10.1093/mnras/263.4.999}, \href
  {https://ui.adsabs.harvard.edu/abs/1993MNRAS.263..999T} {263, 999}

\bibitem[\protect\citeauthoryear{{Tremaine} et~al.,}{{Tremaine}
  et~al.}{2002}]{2002ApJ...574..740T}
{Tremaine} S.,  et~al., 2002, \mn@doi [\apj] {10.1086/341002}, \href
  {https://ui.adsabs.harvard.edu/abs/2002ApJ...574..740T} {574, 740}

\bibitem[\protect\citeauthoryear{{Urry} \& {Padovani}}{{Urry} \&
  {Padovani}}{1995}]{UrryPadovani95}
{Urry} C.~M.,  {Padovani} P.,  1995, \mn@doi [\pasp] {10.1086/133630}, \href
  {https://ui.adsabs.harvard.edu/abs/1995PASP..107..803U} {107, 803}

\bibitem[\protect\citeauthoryear{{Veale}, {Ma}, {Greene}, {Thomas},
  {Blakeslee}, {McConnell}, {Walsh}  \& {Ito}}{{Veale}
  et~al.}{2017}]{2017MNRAS.471.1428V}
{Veale} M.,  {Ma} C.-P.,  {Greene} J.~E.,  {Thomas} J.,  {Blakeslee} J.~P.,
  {McConnell} N.,  {Walsh} J.~L.,   {Ito} J.,  2017, \mn@doi [\mnras]
  {10.1093/mnras/stx1639}, \href
  {https://ui.adsabs.harvard.edu/abs/2017MNRAS.471.1428V} {471, 1428}

\bibitem[\protect\citeauthoryear{{Wada}, {Papadopoulos}  \& {Spaans}}{{Wada}
  et~al.}{2009}]{2009ApJ...702...63W}
{Wada} K.,  {Papadopoulos} P.~P.,   {Spaans} M.,  2009, \mn@doi [\apj]
  {10.1088/0004-637X/702/1/63}, \href
  {https://ui.adsabs.harvard.edu/abs/2009ApJ...702...63W} {702, 63}

\bibitem[\protect\citeauthoryear{{Wagner}, {Bicknell}  \& {Umemura}}{{Wagner}
  et~al.}{2012}]{2012ApJ...757..136W}
{Wagner} A.~Y.,  {Bicknell} G.~V.,   {Umemura} M.,  2012, \mn@doi [\apj]
  {10.1088/0004-637X/757/2/136}, \href
  {https://ui.adsabs.harvard.edu/abs/2012ApJ...757..136W} {757, 136}

\bibitem[\protect\citeauthoryear{{Ward}, {Harrison}, {Costa}  \&
  {Mainieri}}{{Ward} et~al.}{2022}]{2022MNRAS.514.2936W}
{Ward} S.~R.,  {Harrison} C.~M.,  {Costa} T.,   {Mainieri} V.,  2022, \mn@doi
  [\mnras] {10.1093/mnras/stac1219}, \href
  {https://ui.adsabs.harvard.edu/abs/2022MNRAS.514.2936W} {514, 2936}

\bibitem[\protect\citeauthoryear{{W}es {M}c{K}inney}{{W}es
  {M}c{K}inney}{2010}]{mckinney-proc-scipy-2010}
{W}es {M}c{K}inney 2010, in {S}t\'efan van~der {W}alt {J}arrod {M}illman eds,
  {P}roceedings of the 9th {P}ython in {S}cience {C}onference. pp 56 -- 61,
  \mn@doi{10.25080/Majora-92bf1922-00a}

\bibitem[\protect\citeauthoryear{{Wu}, {Feng}  \& {Fan}}{{Wu}
  et~al.}{2018}]{2018ApJ...855...46W}
{Wu} Q.,  {Feng} J.,   {Fan} X.,  2018, \mn@doi [\apj]
  {10.3847/1538-4357/aaac28}, \href
  {https://ui.adsabs.harvard.edu/abs/2018ApJ...855...46W} {855, 46}

\bibitem[\protect\citeauthoryear{{Zhu}, {Zaw}, {Blanton}  \& {Greenhill}}{{Zhu}
  et~al.}{2011}]{2011ApJ...742...73Z}
{Zhu} G.,  {Zaw} I.,  {Blanton} M.~R.,   {Greenhill} L.~J.,  2011, \mn@doi
  [\apj] {10.1088/0004-637X/742/2/73}, \href
  {https://ui.adsabs.harvard.edu/abs/2011ApJ...742...73Z} {742, 73}

\bibitem[\protect\citeauthoryear{pandas~development team}{pandas~development
  team}{2022}]{the_pandas_development_team_2022_7093122}
pandas~development team T.,  2022, pandas-dev/pandas: Pandas,
  \mn@doi{10.5281/zenodo.7093122}, \url
  {https://doi.org/10.5281/zenodo.7093122}

\makeatother
\end{thebibliography}
\bibdata{example.bib}
\bibstyle{mnras}




\appendix
\onecolumn
\section{Emission data and derived values}
\label{Appendix A}
\begin{table*}
\caption{Emission data}
\tiny
\setlength\tabcolsep{1.5pt}
\begin{tabular}{cccccccccccccc}
\hline
Galaxy & {$L_{1.4}$} & {$\Delta L_{1.4}$} & {$E_{\rm 1.4}$} & {$\Delta E_{\rm 1.4}$} & $L_{\rm X,2-10}$ & $\Delta L_{\rm X,2-10}$ & X-ray Source & $L_{\rm mm}$ & $\Delta{L_{\rm mm}}$ & ${\rm log}_{10} \left(\frac{L_{\rm [OIII]}}{\rm erg\,s^{-1}}\right)$ & ${\rm log}_{10}\left(\frac{\Delta L_{\rm [OIII]}}{\rm erg\,s^{-1}}\right)$ & $K_{\rm s}$ & $\Delta K_{\rm s}$ \\
& ($\rm erg\,s^{-1}$) & ($\rm erg\,s^{-1}$) & & & ($\rm erg\,s^{-1}$) & ($\rm erg\,s^{-1}$) &  & ($\rm erg\,s^{-1}$) & ($\rm erg\,s^{-1}$)  & & & (mag) & (mag) \\
(1) & (2) & (3) & (4) & (5) & (6) & (7) & (8) & (9) & (10) & (11) & (12) & (13) & (14)\\
\hline
FRL49 & 9.06E+38 & 9.06E+37 & 0.90 & 0.205 & 1.85E+43 & 1.85E+42 & Chandra & 1.92E+39 & 7.85E+37 & 42.39 & 0.04 & 9.8 & 0.0100 \\
FRL1146 & - & - & - & - & 2.58E+43 & 2.58E+42 & XMM & 2.44E+39 & 1.41E+38 & 40.24 & 0.07 & 11 & 0.00800 \\
MRK567 & 7.48E+38 & 7.91E+37 & 0.29 & 0.205 & - & - & - & 5.84E+38 & - & - & - & 11 & 0.0450 \\
NGC0383 & 3.59E+40 & 3.59E+39 & 3.3 & 0.205 & 3.31E+40 & 3.31E+39 & Chandra & 7.86E+40 & 1.26E+38 & 39.46 & 0.04 & 10 & 0.00700 \\
NGC0404 & 5.36E+34 & 6.99E+33 & 0.49 & 0.227 & 1.58E+37 & 1.58E+36 & Chandra & 9.76E+35 & 3.08E+34 & 37.69 & 0.04 & 10 & 0.00700 \\
NGC0449 & 5.55E+38 & 5.78E+37 & 0.27 & 0.205 & 3.78E+40 & 3.78E+39 & XMM & 7.47E+38 & 3.05E+37 & 41.50 & 0.04 & 12 & 0.0230 \\
NGC0524 & 2.80E+36 & 4.61E+35 & -0.27 & 0.212 & 3.59E+38 & 3.59E+37 & Chandra & 8.81E+38 & 3.56E+36 & 37.55 & 0.04 & 10 & 0.0730 \\
NGC0612 & 1.46E+41 & 1.47E+40 & 3.0 & 0.205 & 7.76E+41 & 7.76E+40 & Chandra & 1.19E+41 & 2.84E+38 & 40.03 & 0.04 & 9.9 & 0.0990 \\
NGC0708 & 3.72E+38 & 3.72E+37 & 1.6 & 0.205 & 2.45E+39 & 2.45E+38 & Chandra & 1.27E+39 & 1.58E+37 & 39.10 & 0.04 & 12 & 0.0600 \\
NGC1194 & 1.18E+37 & 1.18E+36 & 1.5 & 0.871 & 3.50E+41 & 3.50E+40 & XMM & 1.23E+39 & 2.22E+37 & 40.03 & 0.04 & 11 & 0.0320 \\
NGC1387 & 2.64E+36 & 4.22E+35 & -0.18 & 0.212 & 2.14E+39 & 2.14E+38 & Chandra & 1.17E+38 & 6.03E+36 & - & - & 10 & 0.111 \\
NGC1574 & - & - & - & - & - & - & - & 3.52E+38 & 3.50E+36 & - & - & 10 & 0.0980 \\
NGC2110 & 6.34E+38 & 6.76E+37 & - & - & 5.13E+42 & 5.13E+41 & Chandra & 7.72E+39 & 1.64E+38 & 40.41 & 0.04 & 10 & 0.0580 \\
NGC3169 & 5.43E+37 & 5.43E+36 & 0.16 & 0.205 & 3.39E+41 & 3.39E+40 & Chandra & 3.42E+38 & 1.07E+37 & 39.46 & 0.04 & 10 & 0.0920 \\
NGC3351 & 7.08E+36 & 7.08E+35 & - & - & 5.46E+38 & 5.67E+37 & ROSAT & 1.81E+37 & - & 37.28 & 0.15 & 9.0 & 0.0150 \\
NGC3368 & 1.52E+37 & 1.52E+36 & 0.19 & 0.205 & 1.98E+39 & 1.98E+38 & ASCA & 5.40E+37 & - & 38.85 & 0.04 & 9.5 & 0.0680 \\
NGC3607 & 5.67E+36 & 6.57E+35 & 0.010 & 0.206 & 1.45E+39 & 1.45E+38 & Chandra & 3.79E+38 & 2.31E+37 & 39.48 & 0.04 & 9.5 & 0.0420 \\
NGC3862 & 8.13E+40 & 8.13E+39 & - & - & 3.92E+41 & 3.92E+40 & Chandra & 1.52E+41 & 3.43E+39 & 39.20 & 0.04 & 11 & 0.0370 \\
NGC4061 & 5.70E+39 & 5.70E+38 & 3.2 & 0.205 & - & - & - & 5.98E+39 & 4.46E+38 & - & - & 11 & 0.0400 \\
NGC4261 & 3.74E+40 & 4.12E+39 & - & - & 1.17E+41 & 1.17E+40 & Chandra & 6.28E+40 & 4.43E+38 & 39.60 & 0.04 & 9.0 & 0.0150 \\
NGC4429 & 4.59E+34 & 2.30E+35 & -1.8 & 2.18 & 1.31E+39 & 1.31E+38 & Einstein & 1.23E+38 & 9.42E+36 & 38.69 & 0.04 & 10 & 0.0650 \\
NGC4435 & 2.72E+36 & 2.72E+36 & -0.010 & 0.478 & 2.95E+39 & 2.95E+38 & Chandra & 5.69E+37 & 1.92E+36 & 39.02 & 0.04 & 10 & 0.0590 \\
NGC4438 & 2.87E+37 & 2.87E+36 & 0.48 & 0.205 & 1.20E+39 & 1.20E+38 & Chandra & 4.08E+37 & 9.81E+36 & 39.69 & 0.04 & 8.8 & 0.0150 \\
NGC4501 & 9.05E+37 & 9.05E+36 & 0.25 & 0.205 & 1.23E+40 & 1.23E+39 & Chandra & 8.01E+37 & 4.39E+36 & 39.31 & 0.04 & 9.0 & 0.0150 \\
NGC4697 & 1.30E+35 & 1.08E+35 & -1.1 & 0.413 & 3.31E+38 & 3.31E+37 & Chandra & 1.79E+37 & 1.65E+36 & - & - & 9.7 & 0.0660 \\
NGC4826 & 9.30E+36 & 9.03E+35 & 0.40 & 0.204 & 9.23E+37 & 2.05E+37 & Chandra & 8.27E+36 & 1.64E+36 & 39.03 & 0.04 & 8.0 & 0.0150 \\
NGC5064 & 1.93E+38 & 1.93E+37 & 0.89 & 0.205 & - &  - & - & 9.19E+37 & 8.53E+36 & - & - & 11 & 0.0730 \\
NGC5765b & 5.03E+38 & 5.03E+37 & -0.010 & 0.205 & 5.37E+40 & 5.37E+39 & Chandra & 1.15E+39 & 2.17E+38 & - & - & 10.9 & 0.0700 \\
NGC5806 & 1.17E+37 & 1.17E+36 & -0.18 & 0.205 & - & - & - & 1.85E+37 & - & - & - & 11 & 0.0730 \\
NGC5995 & 5.91E+38 & 6.36E+37 & - & - & 2.45E+43 & 2.45E+42 & Chandra & 3.22E+39 & 1.07E+38 & - & - & 10 & 0.0390 \\
NGC6753 & 3.23E+38 & 3.53E+37 & 0.91 & 0.205 & - & - & - & 6.73E+37 & - & - & - & 11 & 0.0990 \\
NGC6958 & 2.34E+37 & 2.81E+36 & 0.67 & 0.207 & - & - & - & 2.92E+39 & 1.51E+37 & - & - & 10 & 0.0600 \\
NGC7052 & 9.21E+38 & 9.30E+37 & 1.8 & 0.205 & 1.05E+40 & 1.05E+39 & Chandra & 1.35E+40 & 6.23E+37 & 39.44 & 0.04 & 11 & 0.0420 \\
NGC7172 & 7.20E+37 & 7.66E+36 & 0.20 & 0.205 & 1.00E+43 & 1.00E+42 & Chandra & 2.74E+39 & 1.04E+38 & - & - & 10 & 0.0560 \\
PGC043387 & -& - & - & - & - & - & - & 8.02E+38 & - & - & - & 10.7 & 0.0110 \\
\hline
\end{tabular}
\parbox[t]{\textwidth}{\textit{Notes:} (1) Galaxy name, (2) 1.4GHz luminosity, (3) 1.4GHz luminosity uncertainty, (4) Excess radio fraction, (5) Excess radio fraction uncertainty, (6) 2-10 keV X-ray luminosity, (7) (2-10) keV X-ray luminosity uncertainty, (8) The telescope used for the X-ray observation, (9) Nuclear mm luminosity, (10) Nuclear mm luminosity uncertainty, (11) [OIII] luminosity, (12) [OIII] luminosity uncertainty, (13) $K_{\rm s}$-band magnitude, (14) $K_{\rm s}$-band magnitude uncertainty.}
\label{tab:A1}
\end{table*}

\begin{table*}
\caption{Derived quantities}
\begin{tabular}{lllll}
\hline
Galaxy & $\dot{M}_{\rm X-ray,acc}$ & $\Delta\dot{M}_{\rm X-ray,acc}$ & $\dot{M}_{\rm [OIII],acc}$ & $\Delta\dot{M}_{[\rm OIII],acc}$ \\
& $(\rm M_{\odot} \, yr^{-1})$ & $(\rm M_{\odot} \, yr^{-1})$ & $(\rm M_{\odot} \, yr^{-1})$ & $(\rm M_{\odot} \, yr^{-1})$\\
(1) & (2) & (3) & (4) & (5) \\
\hline
FRL49 & 5.70E-02 & 5.70E-03 & 1.29E+00 & 1.19E-01 \\
FRL1146 & 8.70E-02 & 8.70E-03 & 9.12E-03 & 1.47E-03 \\
MRK567 & - & - & - & - \\
NGC0383 & 3.55E-05 & 3.55E-06 & 1.51E-03 & 1.39E-04 \\
NGC0404 & 2.50E-08 & 2.50E-09 & 2.57E-05 & 2.37E-06 \\
NGC0449 & 4.10E-05 & 4.10E-06 & 1.66E-01 & 1.53E-02 \\
NGC0524 & 3.98E-07 & 3.98E-08 & 1.86E-05 & 1.72E-06 \\
NGC0612 & 1.19E-03 & 1.19E-04 & 5.63E-03 & 5.18E-04 \\
NGC0708 & 2.48E-06 & 2.48E-07 & 6.61E-04 & 6.09E-05 \\
NGC1194 & 4.77E-04 & 4.77E-05 & 5.63E-03 & 5.18E-04 \\
NGC1387 & 2.17E-06 & 2.17E-07 & - & - \\
NGC1574 & - & - & - & -\\
NGC2110 & 1.15E-2 & 1.15E-3 & 1.91E-02 & 1.76E-03 \\
NGC3169 & 4.60E-04 & 4.60E-05 & 1.51E-03 & 1.39E-04 \\
NGC3351 & 5.88E-07 & 6.10E-08 & 1.00E-05 & 3.46E-06 \\
NGC3368 & 2.01E-06 & 2.01E-07 & 3.72E-04 & 3.42E-05 \\
NGC3607 & 1.49E-06 & 1.49E-07 & 1.59E-03 & 1.46E-04 \\
NGC3862 & 5.43E-04 & 5.43E-05 & 8.32E-04 & 7.67E-05 \\
NGC4061 & - & - & - & - \\
NGC4261 & 1.39E-04 & 1.39E-05 & 2.09E-03 & 1.93E-04 \\
NGC4429 & 1.34E-06 & 1.34E-07 & 2.57E-04 & 2.37E-05 \\
NGC4435 & 2.98E-06 & 2.98E-07 & 5.50E-04 & 5.06E-05 \\
NGC4438 & 1.24E-06 & 1.24E-07 & 2.57E-03 & 2.37E-04 \\
NGC4501 & 1.26E-05 & 1.26E-06 & 1.07E-03 & 9.87E-05 \\
NGC4697 & 3.68E-07 & 3.68E-08 & - & - \\
NGC4826 & 1.16E-07 & 2.56E-08 & 5.63E-04 & 5.18E-05 \\
NGC5064 & - & - & - & - \\
NGC5765b & 5.96E-05 & 5.96E-06 & - & - \\
NGC5806 & - & - & - & - \\
NGC5995 & 8.14E-02 & 8.14E-03 & - & - \\
NGC6753 & - & - & - & - \\
NGC6958 & - & - & - & - \\
NGC7052 & 1.07E-05 & 1.07E-6 & 1.45E-03 & 1.33E-04 \\
NGC7172 & 2.63E-02 & 2.63E-03 & - & - \\
PGC043387 & - & - & - & - \\
\hline
\end{tabular}
\label{tab:A2}
\parbox[t]{\textwidth}{\textit{Notes:} (1) Galaxy name. (2) X-ray-traced accretion rate. (3) X-ray-traced accretion rate uncertainty. (4) [OIII]-traced accretion rate. (5) [OIII]-traced accretion rate uncertainty.}
\end{table*}

\begin{table*}
\caption{CO integrated intensity data}
\begin{tabular}{lllllllll}
\hline
Galaxy & $I_{\rm CO}$ & $\Delta I_{\rm CO}$ & $I_{\rm CO}$ & $\Delta I_{\rm CO}$ & $I_{\rm CO}$ & $\Delta I_{\rm CO}$ & $I_{\rm CO}$ & $ \Delta I_{\rm CO}$\\
& \multicolumn{2}{c}{(200\,pc)} & \multicolumn{2}{c}{(100\,pc)} & \multicolumn{2}{c}{(75\,pc)} & \multicolumn{2}{c}{(50\,pc)} \\
& $\rm (Jy \, km \, s^{-1})$ & $(\rm Jy \, km \, s^{-1})$ & $\rm (Jy \, km \, s^{-1})$ & $(\rm Jy \, km \, s^{-1})$ & $(\rm Jy \, km \, s^{-1})$ & $( \rm Jy \, km \, s^{-1})$ & $(\rm Jy \, km \, s^{-1})$ & $(\rm Jy \, km \, s^{-1})$\\
(1) & (2) & (3) & (4) & (5) & (6) & (7) & (8) & (9)\\
\hline
FRL49 & 7.75 & 0.78 & 2.39 & 0.24 & - & - & - & - \\
FRL1146 & - & - & - & - & - & - & - & - \\
MRK567 & 9.58 & 0.96 & 3.98 & 0.4 & - & - & - & - \\
NGC0383 & 7.60 & 0.76 & 2.74 & 0.27 & 1.6 & 0.16 & 0.64 & 0.06 \\
NGC0404 & 39.62 & 3.96 & 39.73 & 3.97 & 39.07 & 3.91 & 36.14 & 3.61 \\
NGC0449 & - & - & - & - & - & - & - & - \\
NGC0524 & 11.64 & 1.16 & 3.77 & 0.38 & 2.48 & 0.25 & 1.29 & 0.13 \\
NGC0612 & 2.22 & 0.22 & 0.58 & 0.06 & 0.44 & 0.05 & 0.32 & 0.04 \\
NGC0708 & 9.24 & 0.92 & 3.18 & 0.32 & 2.05 & 0.21 & 0.95 & 0.1 \\
NGC1194 & 2.58 & 0.26 & 1.13 & 0.11 & 0.76 & 0.08 & - & - \\
NGC1387 & 27.95 & 2.80 & 6.8 & 0.68 & 3.78 & 0.38 & 1.62 & 0.16 \\
NGC1574 & 3.98 & 0.40 & 3.53 & 0.35 & 2.89 & 0.29 & 1.6 & 0.16 \\
NGC2110 & 4.62 & 0.46 & 1.21 & 0.12 & - & - & - & - \\
NGC3169 & 125.84 & 12.58 & 43.18 & 4.32 & 26.99 & 2.7 & - & - \\
NGC3351 & 104.99 & 10.50 & 55.74 & 5.57 & 32.78 & 3.28 & 18.57 & 1.86 \\
NGC3368 & 170.74 & 17.07 & 54.79 & 5.48 & 35.54 & 3.55 & 19.23 & 1.92 \\
NGC3607 & 37.39 & 3.74 & 15.56 & 1.56 & 10.37 & 1.04 & - & -  \\
NGC3862 &  - & - & - & - & - & - & - & - \\
NGC4061 & 1.21 & 0.12 & 0.34 & 0.03 & 0.21 & 0.02 & - & -\\
NGC4261 & 5.91 & 0.59 & 5.49 & 0.55 & 4.59 & 0.46 & 2.98 & 0.3 \\
NGC4429 & 17.38 & 1.74 & 3.64 & 0.36 & 2 & 0.2 & 0.64 & 0.06 \\
NGC4435 & 35.67 & 3.57 & 15.68 & 1.57 & 9.91 & 0.99 & 5.28 & 0.53 \\
NGC4438 & 134.20 & 13.42 & 52.63 & 5.26 & 32.86 & 3.29 & 15.26 & 1.53 \\
NGC4501 & 137.79 & 13.78 & 63.66 & 6.37 & 40.61 & 4.06 & 20.26 & 2.03 \\
NGC4697 & 1.90 & 0.19 & 1.55 & 0.16 & 1.23 & 0.12 & 0.74 & 0.07 \\
NGC4826 & 276.92 & 27.69 & 193.02 & 19.3 & 148.73 & 14.87 & 74.96 & 7.5 \\
NGC5064 & 27.91 & 2.79 & 9.7 & 0.97 & 5.27 & 0.53 & 2.58 & 0.26 \\
NGC5765b & 4.70 & 0.47 & - & - & - & - & - &  -\\
NGC5806 & 19.20 & 1.92 & 7.53 & 0.75 & 5.46 & 0.55 & 3.09 & 0.31 \\
NGC5995 & - & - & - & - & - & - & - & - \\
NGC6753 & 57.58 & 5.76 & 19.66 & 1.97 & 11.8 & 1.18 & 5.65 & 0.56 \\
NGC6958 & 12.76 & 1.28 & 3.76 & 0.38 & 2.3 & 0.23 & 1.03 & 0.11 \\
NGC7052 & 5.06 & 0.51 & 2.2 & 0.22 & 1.33 & 0.13 & 0.55 & 0.06 \\
NGC7172 & 47.03 & 4.70 & 6.19 & 0.62 & 2.85 & 0.29 & 1.14 & 0.11 \\
PGC043387 & - & - & - & - & - & - & - & -\\
\hline
\end{tabular}
\label{tab:A3}
\parbox[t]{\textwidth}{\textit{Notes:} (1) Galaxy name. (2) 200pc radius aperture integrated CO intensity. (3) 200pc aperture CO intensity uncertainty. (4)-(9) follows the same trend for the 100, 75 and 50\,pc radius apertures.}
\end{table*}

\begin{table*}
\begin{tabular}{ccccc}
\hline
\multicolumn{5}{c}{Median values for samples}\\
\hline
& This Work & Garcia-Burillo+21 & Izumi+16 & Babyk+19 \\
\hline
log~$\rm L_{bol}(erg~s^{-1})$ & 43.2 & 43.0 & 43.3 & -   \\
log~$\rm M_{BH}(M_{\odot})$ & 8.04 & 7.09 & 7.38 & - \\
log~($\lambda_{\rm edd}$) & -2.57 & -2.20 & -2.17 & - \\
log~$\rm P_{jet}(erg~s^{-1})$ & 42.3 & - & - & 42.3 \\
\hline
\multicolumn{5}{c}{KS test}\\
\hline
\multicolumn{3}{c}{Samples} & \multicolumn{2}{c}{\textit{p}-values}\\
\hline
\multicolumn{3}{c}{This Work-GB+21 ($\rm L_{bol}$)} & \multicolumn{2}{c}{0.015}\\
\multicolumn{3}{c}{This Work-Izumi+16 ($\rm L_{bol}$)} & \multicolumn{2}{c}{0.001}\\
\multicolumn{3}{c}{This Work-GB+21 ($\rm M_{BH}$)} & \multicolumn{2}{c}{0.0003}\\
\multicolumn{3}{c}{This Work-Izumi+16 ($\rm M_{BH}$)} & \multicolumn{2}{c}{7.27$\times10^{-5}$}\\
\multicolumn{3}{c}{This Work-GB+21 ($\rm \lambda_{edd}$)} & \multicolumn{2}{c}{0.03}\\
\multicolumn{3}{c}{This Work-Izumi+16 ($\rm \lambda_{edd}$)} & \multicolumn{2}{c}{0.003}\\
\multicolumn{3}{c}{This Work-Babyk+19 ($\rm P_{jet}$)} & \multicolumn{2}{c}{0.53}\\
\hline
\end{tabular}
\caption{Median and KS test \textit{p}-values for the samples used in this work}
\label{tab:stats}
\end{table*}

\begin{figure}
    \centering
    \includegraphics[width=9cm]{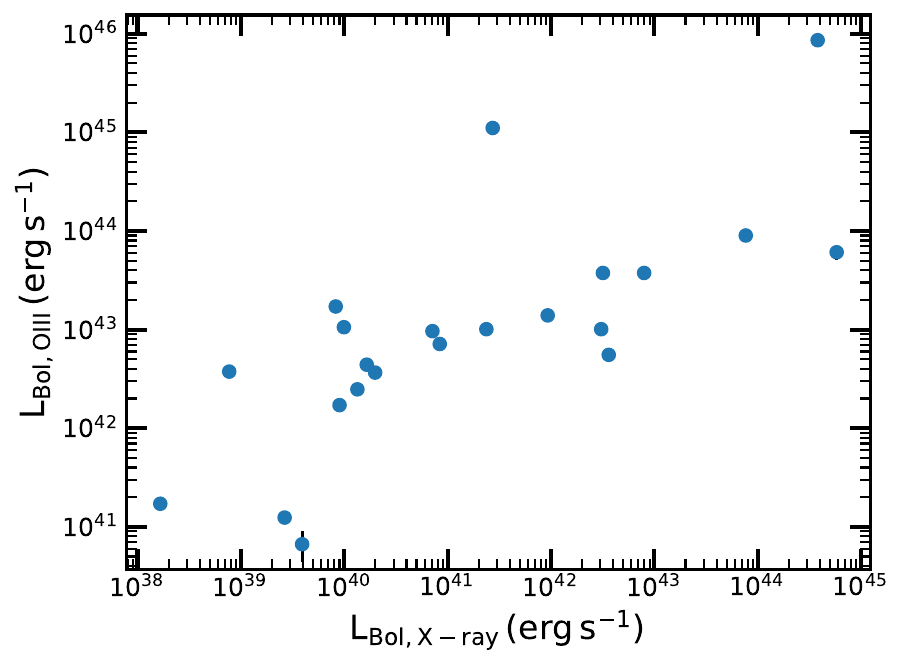}
    \caption{$L_{\rm Bol}$ derived from X-ray versus $L_{\rm Bol}$ derived from [OIII]}
    \label{fig:X-ray-OIII}
\end{figure}

\bsp	
\label{lastpage}
\end{document}